\documentclass[twocolumn,amsmath,amssymb,prc,superscriptaddress,floatfix,showpacs,nofootinbib]{revtex4}

\usepackage{graphicx}
\usepackage{dcolumn}
\usepackage{bm}
\usepackage[utf8]{inputenc}
\usepackage{amssymb}
\usepackage{amsmath}
\usepackage{bbold}
\usepackage{pstricks}
\usepackage{color}
\usepackage{slashed}
\usepackage{braket}
\usepackage{hyperref}
\allowdisplaybreaks

\newcommand{\bracket}[1]{\left\langle#1\right\rangle}

\newcommand*\xbar[1]{%
  \hbox{%
    \vbox{%
      \hrule height 0.5pt 
      \kern0.5ex
      \hbox{%
        \kern-0.1em
        \ensuremath{#1}%
        \kern-0.1em
      }%
    }%
  }%
} 

\graphicspath{
{./}
{./Figures/}
}

\begin{document}

\title{Fermion production from real-time lattice gauge theory\\ in the classical-statistical regime}

\author{V. Kasper}
\email{v.kasper@thphys.uni-heidelberg.de}
\affiliation{Institut f\"{u}r Theoretische Physik, Universit\"{a}t Heidelberg,
  Philosophenweg 16, 69120 Heidelberg, Germany}

\author{F. Hebenstreit}
\affiliation{Institut f\"{u}r Theoretische Physik, Universit\"{a}t Heidelberg,
  Philosophenweg 16, 69120 Heidelberg, Germany}

\author{J. Berges}
\affiliation{Institut f\"{u}r Theoretische Physik, Universit\"{a}t Heidelberg,
  Philosophenweg 16, 69120 Heidelberg, Germany}
\affiliation{ExtreMe Matter Institute EMMI, GSI Helmholtzzentrum,
  Planckstra\ss e 1, 64291 Darmstadt, Germany}

\begin{abstract}
We investigate the real-time dynamics of U(1) and SU(N) gauge theories coupled to fermions on a lattice. 
While real-time lattice gauge theory is not amenable to standard importance sampling techniques, for a large class of time-dependent problems the quantum dynamics can be accurately mapped onto a classical-statistical ensemble. 
We illustrate the genuine quantum contributions included in this description by giving a diagrammatic representation in a series expansion. 
The non-perturbative simulation method is then applied to electron-positron production in quantum electrodynamics in three spatial dimensions.  
We compare to analytic results for constant background field and demonstrate the importance of back-reaction of the produced fermion pairs on the gauge fields.        
\end{abstract}

\maketitle


\section{Introduction}

The quantum dynamics of strong gauge fields coupled to fermions provides an important challenge for theoretical developments. 
A most prominent application concerns the Schwinger mechanism~\cite{Sauter:1931zz,Heisenberg:1935qt,Schwinger:1951nm}, where charged particles such as electron-positron pairs are produced spontaneously from an applied electric field. 
This non-perturbative quantum phenomenon is exponentially suppressed unless the field strength exceeds a critical size, which is given by an electric field $E \sim m^2/g \sim 10^{18} V/m$ for quantum electrodynamics (QED) with electron mass $m$ and gauge coupling $g$. 
The subject of pair production is also relevant in the context of quantum chromodynamics (QCD). 
In the first stages of nuclear collisions at high energy, strong color fields are expected to play an important role for the space-time evolution of the Quark Gluon Plasma. 
In the Color Glass Condensate framework, the characteristic field strengths are of order $1/g$ in the running gauge coupling $g$~\cite{Gelis:2010nm}. 
Correspondingly, the probability for quark pair production is not suppressed. 
The investigation of similar phenomena for strong fields in scalar theories coupled to fermions has also pointed out the dramatic role of fluctuations for the production of fermions~\cite{Berges:2010zv}. 

Since identical fermions cannot occupy the same state, their quantum nature is highly relevant and a consistent quantum treatment of the fermion dynamics is of crucial importance. 
In thermal equilibrium the quantum theory can be mapped onto a statistical mechanics problem, where the time variable is analytically continued to imaginary values. 
This statistical theory can then be simulated on a lattice by importance sampling techniques. 
However, fermion production from strong fields or fluctuations represents an out-of-equilibrium problem that requires the description of the real-time evolution. 
Nonequilibrium problems are not amenable to an Euclidean formulation and for real times standard importance sampling is not possible because of a non-positive definite probability measure. 
Though there is no general solution for this problem yet, it is important to note that for a large class of time-dependent problems the quantum dynamics may be accurately mapped onto a classical-statistical ensemble which can be simulated on a lattice~\cite{Son:1996zs,Khlebnikov:1996mc,Prokopec:1996rr,Aarts:1998td,Gelis:2013oca}. 
This agreement has been explicitly demonstrated for scalar quantum field theories~\cite{Aarts:2001yn,Polkovnikov:2003,Arrizabalaga:2004iw,Berges:2007ym,Berges:2008wm} coupled to fermions~\cite{Berges:2010zv,Berges:2013oba}, where direct comparisons to results from sophisticated resummation techniques for the nonequilibrium quantum theory are available and the range of validity of the classical approach can be verified \cite{Berges:2013lsa,Epelbaum:2014yja}. 
Derivations of classical descriptions for the underlying quantum dynamics in scalar and pure gauge field theories have also been given in the context of kinetic theory~\cite{Mueller:2002gd, Jeon:2004dh, Mathieu:2014aba}. 
In its range of validity the relation between kinetic theory and classical-statistical lattice simulations for gauge theories has been studied in detail~\cite{Berges:2013eia,Berges:2013fga,York:2014wja}.    

In this work, we investigate the real-time dynamics of U(1) and SU(N) gauge theories coupled to fermions on a lattice. 
Starting from the functional integral of the quantum theory, we derive its classical-statistical approximation and discuss its range of validity. 
The genuine quantum contributions included in this description are illustrated by giving a diagrammatic representation in a series expansion. 
We then apply this non-perturbative simulation method for the first time to electron-positron production in QED in three spatial dimensions, which extends previous one-dimensional results~\cite{Hebenstreit:2013qxa}.
We compare to analytic expressions for constant background field and demonstrate the importance of back-reaction of the produced fermion pairs on the gauge fields.        

This paper is organized as follows.
In Sec.~\ref{sec:csft} we derive the classical-statistical approximation for Abelian and non-Abelian gauge theories coupled to fermions. 
Sec.~\ref{sec:diag} provides a diagrammatic interpretation. 
The nonperturbative simulation procedure on a Minkowskian lattice is described in Sec.~\ref{sec:rtsim}.
We present the results concerning electron-positron production in QED in Sec.~\ref{sec:seff}. 
Conclusions and an outlook are given in Sec.~\ref{sec:conc}.

\section{Real-time lattice gauge theory}
\label{sec:csft}

\begin{figure}[t]
 \includegraphics[width=0.95\columnwidth]{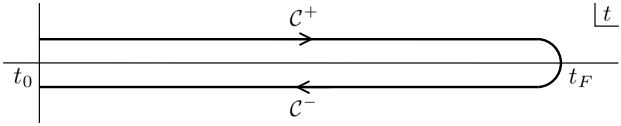}
 \caption{\label{fig:SK} Graphical representation of the real-time Schwinger-Keldysh contour where $\mathcal{C}_+$ denotes the forward branch and $\mathcal{C}_-$ is the backward branch.}
\end{figure}

Real-time quantum field theory can be formulated in terms of a functional integral on the Schwinger-Keldysh contour, as displayed in Fig.~\ref{fig:SK}, which starts at some initial time $t_0$ running along the real-time axis in the forward direction and then backwards~\cite{Schwinger:1961,Keldysh:1965}. 
To be specific we discretize space-time on a hypercubic lattice
\begin{align}
\Lambda = \{(n_0,\mathbf{n}) \,|\,  &n_0 \in 0, \dots , 2 N_T\, ;\, n_i  \in  0, \dots , N_i-1 \} ,
\end{align}
with $\mathbf{n}=(n_1,n_2,n_3)$ and $i\in\{1,2,3\}$.
In the following, a vector $\hat{\mu}$ points along the time contour for $\mu=0$ and along the direction of the spatial coordinate axes for $\mu\in\{1,2,3\}$, where we consider an isotropic spatial lattice with spacings $a_i=a_S$ and the number of lattice points $N_i=N_S$ for $i\in\{1,2,3\}$.
The discretization of the time contour is 
\begin{align}
t_\mathcal{C}(n_0) = 
\begin{cases}
t_0 + a_T \, n_0  ,& 0 \leq n_0 \leq N_T  \\
t_F - a_T \, (n_0-N_T)  ,& N_T +1 \leq n_0 \leq 2 N_T
\end{cases} 
\end{align}
\noindent 
where the final time $t_F=t_0+a_TN_T$ is chosen as large as necessary
and $n_0$ is a non-negative integer. The lattice spacing in the temporal direction is thus given by
\begin{align}
a_0 \equiv  \operatorname{sgn}_\mathcal{C}a_T \, ,
\end{align} 
where $\operatorname{sgn}_\mathcal{C}$ is $+1$ on the forward branch ($0\leq n_0\leq N_T$) and $-1$ on the backward branch ($N_T +1 \leq n_0 \leq 2 N_T$) of the Schwinger-Keldysh contour.

For the pure gauge part of the considered gauge theories, we employ the Wilsonian action on a Minkowskian lattice~\cite{Ambjorn:1990pu,Berges:2007re}
\begin{align}
 \label{eq:gaugeaction}
 S_G[U]&=\frac{2}{g_0^2}\sum\limits_{n \in \Lambda}{\frac{a_S}{a_0}}\sum_{i}{\operatorname{Re}\mathrm{tr}\left( {\mathbb{1} - {U_{0i,n}}} \right)} \notag \\
  &-\frac{2}{g_S^2}\sum\limits_{n \in \Lambda}{\frac{a_0}{a_S}}\sum_{\substack{i,j\\ i<j}}{\operatorname{Re}\mathrm{tr}\left( {\mathbb{1} - {U_{ij,n}}} \right)}  \, ,
\end{align}
which is expressed in terms of the gauge-invariant plaquettes
\begin{align}
 U_{\mu \nu ,n} = U_{\mu ,n}U_{\nu ,n + \hat \mu }U_{\mu ,n + \hat \nu }^\dagger U_{\nu ,n}^\dagger \, .
\end{align}
\newline
\noindent The link variables $U_{\mu,n}$ are located between lattice sites $n$ and $n+\hat{\mu}$ and point in the direction of $\hat{\mu}$.
For link variables pointing in the direction of $-\hat{\mu}$, we use the definition
\begin{align}
U_{-\mu,n} = U^{\dagger}_{\mu,n-\hat{\mu}} \, .
\end{align}

In the following, we will assume $g_0=g_S=g$ for simplicity. 
The prefactors in $S_G$ have been chosen such that the correct naive continuum limit is obtained when using the standard normalization $\operatorname{tr} \,  [t^{a}, t^{b}] = 1/2  \, \delta_{ab}$ of the SU(N) generators $t^a$, with the adjoint gauge index  $a \in \{ 1 , \dots, N^2 -1  \}$.
It has to be noted, however, that the overall factor $2$ has to be replaced by $1$ for the case of a U(1) gauge theory.

In the fermion sector, we consider first the naive discretization for simplicity and defer the issue of fermion doublers to Sec.~\ref{sec:rtsim}: 
\begin{widetext}
\begin{align}
\label{eq:fermionaction}
S_F[\bar{\psi}, \psi ,U] = S_{if} + \frac{i}{2}\sum_{\substack{n\in\Lambda\\ n_0\neq 0 ,2N_T  }} {\frac{a_0a_S^3}{a_\mu}{\bar{\psi}_n \gamma^\mu\left[ U_{\mu,n}\psi_{n+\hat{\mu}} - U_{-\mu,n}\psi_{n-\hat{\mu}}\right]} }  - m\sum\limits_{n \in \Lambda } {a_0a_S^3\bar{\psi}_n\psi_n}  \, .
\end{align}
\end{widetext}
We choose the central difference prescription for the fermions and $S_{if}$ contains the initial and final discretization of the fermions:
\begin{align}
S_{if} &=   i {\frac{a_0a_S^3}{a_\mu}{\bar{\psi}_0 \gamma^\mu\left[ U_{\mu,0}\psi_{0+\hat{\mu}} - \psi_{0}\right]} }  \notag \\
&+ i {\frac{a_0a_S^3}{a_\mu}{\bar{\psi}_{2N_T} \gamma^\mu\left[ \psi_{2N_T} - U_{-\mu,2N_T} \psi_{2N_T-1}\right]} } 
\end{align}
For a given initial density matrix $\rho(t_0)$, which may describe thermal equilibrium or nonequilibrium, expectation values of observables $O(U,\bar{\psi},\psi)$ can be obtained from the functional integral
\begin{align} 
 \label{eq:observables} 
 \langle O(U,\bar{\psi},\psi) \rangle = &\int {\left[ {dU} \right]} \int {\left[ {d\bar \psi \, d\psi } \right]}\,\rho(t_0)\, \notag \\ & \times O(U,\bar{\psi},\psi)\, \exp(iS_G + iS_F) \ ,
\end{align}
with classical action $S_G + S_F$ for the gauge and fermion degrees of freedom. Here we use the abbreviation
\begin{subequations}
\begin{align}
\int {\left[ {dU} \right]} &= \prod\limits_{\substack{n\in\Lambda\\ \mu}} {\int {d{U_{\mu,n}}} } \ , 
\label{eq:Haar} \\
\int {\left[ {d\bar \psi d\psi } \right]} &= \prod\limits_{n\in\Lambda} {\int {d{{\bar \psi}_n}d{\psi _n}} } \ .
\end{align}
\end{subequations}
The initial density matrix $\rho(t_0)$ depends on $U_{\mu,n}$, $\bar{\psi}_n$ and $\psi_n$ only at $n_0\in\{0,2N_T\}$.

We emphasize that the real-time 'partition function' on the closed time path,
\begin{align} 
 \label{ZEqnNum201414} 
 {Z_\mathcal{C}} = \int {\left[ {dU} \right]} \int {\left[ {d\bar \psi \, d\psi } \right]}\,\rho(t_0)\, \exp(iS_G + iS_F) \, ,
\end{align} 
is normalized to one since the contributions from the forward and backward branch of the time contour cancel in the absence of source terms: $Z_\mathcal{C} = 1$. 
This straightforwardly takes the normalization for the computation of observables (\ref{eq:observables}) into account. 
A generating functional for correlation functions can be obtained from (\ref{ZEqnNum201414}) by introducing source terms.\footnote{For an introduction to nonequilibrium generating functionals see Ref.~\cite{Berges:2004yj}.} 
However, for the following discussion it will be convenient to disregard possible source terms for notational simplicity. 
We perform all manipulations on the integrand of (\ref{ZEqnNum201414}) keeping in mind that we can always introduce sources, or directly insert the corresponding observables, to get the respective expectation values.  

The complex factor $\sim \exp(iS_G + iS_F)$ appearing in the full partition function $(\ref{ZEqnNum201414})$ preempts the use of standard importance sampling techniques. In the following, we describe an alternative method for an approximate estimate and discuss its range of validity in Sec.~\ref{sec:diag}.\footnote{Although we consider only quadratic fermionic actions, we note that similar techniques can also be applied to theories with non-Gaussian fermion interaction terms at the expense of introducing composite fields via a Hubbard-Stratonovich transformation.}

\subsection{U(1) gauge theory}
\label{sec:csftU1}

In this section we derive the classical-statistical approximation for QED.
While all simulations will be performed directly in terms of the link variables as described in Sec.~\ref{sec:rtsim}, for the following discussion it is convenient to parametrize the link variables $U_{\mu,n}$ in terms of the gauge fields $A_{\mu,n}$:\footnote{In principle, the parametrization of the Haar measure in terms of gauge fields introduces non-trivial contributions to the functional integral \cite{Kawai:1980ja}.
We will further discuss this issue in Sec.~\ref{sec:sun} on SU(N) gauge theories.
For the U(1) gauge theory, however, this term is trivial and does not modify the functional integral.}
\begin{align}
U_{\mu,n} = \exp \left( i g a_{\mu} A_{\mu,n} \right) \,.
\end{align}
Accordingly, the partition function reads
\begin{align} 
 \label{eq:GenFuncU1}
 {Z_\mathcal{C}} = \int {\left[ {dA} \right]} \int {\left[ {d\bar \psi \, d\psi } \right]}\,\rho(t_0)\exp\left(iS_G+iS_F\right) \ .
\end{align} 

In the following, we assume the initial density matrix to be quadratic in the fermion fields:\footnote{Correlated initial states can be implemented in this case by introducing an imaginary time branch~\cite{Danielewicz:1982ca}, which we will not consider here.}
\begin{align}
 \label{eq:density}
\rho(t_0) = \exp \left[- \sum _{n,m\in\Lambda}a_0 a_S^3a_0 a_S^3 \bar{\psi}_n(\mathcal{K}^{-1})_{nm}\psi_m \right] \rho_G(A)   \, ,
\end{align}
where $\mathcal{K}^{-1}$ has only support at $n_0, m_0 \in \{0, 2N_T\}$ and may also depend on the initial gauge fields. 
Accordingly, the pure gauge part $\rho_G(A)$ depends only on the initial field configuration $A_{\mu,n}$ with $n_0 \in \{0, 2N_T\}$. 
This guarantees that the fermionic fields appear at most quadratically in the exponent of the functional integral \eqref{eq:GenFuncU1}, which can be written as an effective fermion action
\begin{align}
 S_{FQ}[A] = \sum\limits_{n,m \in \Lambda} \, {{{\bar \psi }_n}}{i\Delta_{\mathcal{C}}[A]^{-1} _{nm}}{\psi _m} \, .
\end{align}
Here, $ i\Delta_{\mathcal{C}}[A]^{-1}$ denotes the inverse fermion propagator on the Schwinger-Keldysh contour which depends on the gauge field $A$. 
Due to the fact that the Grassmann variables appear at most quadratically in the exponent, we can integrate them out, resulting in
\begin{align} 
 \label{eq:IntegratingOutFermions}
 {Z_\mathcal{C}} = \int {\left[ {dA} \right]} \,\rho_G(A)\, \exp \left( \text{Tr} \log \Delta_{\mathcal{C}}[A]^{-1} + i{S_G} \right) \,.
\end{align}
The term $\operatorname{Tr}\log \Delta_{\mathcal{C}}[A]^{-1}$, where the trace involves Dirac as well as space-time indices on the contour, corresponds to a non-local effective interaction mediated by the fermionic degrees of freedom.

To proceed, we label the gauge fields on the forward/backward branch of the Schwinger-Keldysh contour by $+$/$-$, respectively, and write:
\begin{subequations}
\begin{align}
 A_{\mu,n}^{+} &= \bar{A}_{\mu,n} +\frac{1}{2} \tilde{A}_{\mu,n}  \, ,  \\
 A_{\mu,n}^{-} &= \bar{A}_{\mu,\bar{n}} -\frac{1}{2} \tilde{A}_{\mu,\bar{n}} \, ,
\end{align}
\end{subequations}
with $\bar{n} = (2N_T-1 -n_0,\mathbf n)$. 
In the following $\bar{A}$ and $\tilde{A}$ are denoted as 'classical' and 'quantum' fields, respectively. 
In terms of these new variables, the partition function takes the form
\begin{align} 
 \label{eq:GenFuncPureBosonic}
 {Z_\mathcal{C}} = \int {\left[ {d \bar{A}} \right][d\tilde{A}]\,  \rho_G(A)\exp \left( \operatorname{Tr} \log \Delta_{\mathcal{C}}[A]^{-1}+ i{S_G[A]} \right)} \, ,
\end{align}
with the notation $A=\bar{A}+\frac{1}{2}\operatorname{sgn}_\mathcal{C}\tilde{A}$.

The gauge action term $S_G[A]$ can be written as
\begin{align}
  S_G[A]=-\frac{1}{4}\sum_{n\in\Lambda}a_{0}a_S^3 F_{\mu\nu,n}[A] F^{\mu\nu}_n[A] \ ,
\end{align}
up to higher orders in the lattice spacing and $F^{\mu\nu}_n$ denotes the field strength tensor
\begin{align}
 F_{\mu\nu,n}[A] = \partial_{\mu} A_{\nu,n} - \partial_{\nu} A_{\mu,n} \, .
\end{align}
Writing this action in the basis of $\bar{A}$ and $\tilde{A}$,
\begin{align} 
 \label{eq:TildeAction}
 {S_G}[ \bar{A}, \tilde{A} ] =  a_{T}a_S^3\sum\limits_{n \in \Lambda^{+}} {\tilde A}_{\nu,n }\partial_\mu F^{\mu\nu}_{n} [\bar{A}]\, ,
\end{align}
one observes that it is linear in $\tilde{A}$. Here $\Lambda^{+}$ denotes the forward branch of the Schwinger-Keldysh contour and $F^{\mu\nu}_n[\bar{A}]$ is the field strength tensor of the classical field.

Expanding also the term $\operatorname{Tr}\log\Delta_{\mathcal{C}}[A]^{-1}$ appearing in the functional integral (\ref{eq:GenFuncPureBosonic}) to linear order in the quantum field $\tilde{A}$, we obtain
\begin{align}  
 \label{eq:expansion}
 \operatorname{Tr}\log \Delta_{\mathcal{C}}^{-1} [A]=&\operatorname{Tr} \log \Delta_{\mathcal{C}}^{-1}[\bar{A}] \notag \\ &+\frac{ig}{2} \mathrm{Tr} \{\Delta_{\mathcal{C}}[\bar{A}] \operatorname{sgn}_\mathcal{C} \slashed {\tilde {A}}\} + \ldots \, ,
\end{align} 
where $\slashed {\tilde {A}} \equiv \gamma^\mu \tilde{A}_\mu$ and $\Delta_{\mathcal{C}}[\bar{A}] $ is the fermion propagator in the background of the classical field $\bar{A}$ with $\tilde{A} = 0$.
In the continuum the leading term of the expansion \eqref{eq:expansion} does not contribute to the functional integral: 
Since no $\tilde{A}$ appears in this expression, the contribution from the forward branch is canceled by the contribution from the backward branch:
\begin{align} 
 \label{eq:normalization}
 e^{\operatorname{Tr} \log  \Delta_{\mathcal{C}}^{-1}[\bar{A}] }= \det  \Delta_{\mathcal{C}}^{-1}[\bar{A}]= 1 \, .
\end{align}
Hence, we set this term to one in the following discussion. The linear order in $\tilde{A}$, on the other hand, is given by
\begin{align}
 \label{eq:currentcoupling}
 &\frac{ig}{2}\operatorname{Tr}\{\Delta_{\mathcal{C}}[\bar{A}] \operatorname{sgn}_\mathcal{C} \slashed {\tilde {A}}\}=-ia_Ta_S^3\sum_{n\in\Lambda^+}\bar{j}^{\nu}_n\tilde{A}_{\nu,n} \ ,
\end{align}
where we introduced the fermion current
\begin{align}
\bar{j}_{n}^{\nu} = \frac{g}{2} \operatorname{tr} \{ \braket{[\bar{\psi}_n,\psi_n]}_{\bar{A}} \gamma^{\nu} \} \, . 
\end{align}
Here  $\operatorname{tr}$ is the trace over the Dirac indices and $\braket{[\psi_n,\bar{\psi_n}]}_{\bar{A}}$ is the commutator expectation value in the presence of the classical field $\bar{A}$. 
For a detailed derivation of this expression and how to determine $\braket{[\psi_n,\bar{\psi_n}]}_{\bar{A}}$ we refer to Appendix~\ref{app:A}. 
 
As a consequence, the partition function with an expansion of $\operatorname{Tr}\log\Delta_{\mathcal{C}}^{-1}[A]$ to linear order in the quantum field $\tilde{A}$, is given by
\begin{align}
 Z_\mathcal{C}^{\rm cl} = &\int {[ {d\bar A} ][ {d\tilde A} ]} \,\rho_G(A)\, \nonumber \\
 &\times \exp\Big\{ia_Ta_S\sum_{n\in\Lambda^+}{\tilde{A}_{\nu,n}\left(\partial_\mu F^{\mu\nu}_n[\bar{A}] - \bar{j}^\nu_n \right)}\Big\} \,.
\end{align}
This approximate expression, whose physical interpretation will be explained in more detail in Sec.~\ref{sec:diag}, establishes the classical-statistical description.  In order to see how this functional integral can be evaluated using standard techniques, we integrate out the quantum field $\tilde{A}$. To this end, one performs a Fourier transformation of $\rho_G(A)$ with respect to $\tilde{A}$, resulting in the Wigner transform:
\footnote{Changing variables introduces a Jacobian in the functional integral, which can be taken to be constant in the following discussion~\cite{Jeon:2004dh}.}
\begin{align} 
 \label{eq:Wigner}
 \rho _{G} (A) =\int d\Pi _\bold{0} \, {\rho _{W}}\left( {{\bar A}_{\bold{0}}},{\Pi_{\bold{0}}} \right) \exp \left( i \sum_{\bold n}a_S^3 \, \Pi_\bold{0}^{\mu}\tilde{A}_{\mu,\bold{0}} \right) \ ,
\end{align}
where we use the notation $\bold {0 }= (0, \bold{n})$.
We note that $\Pi_\bold{0}$ is the conjugate variable of the classical field $\bar{A}_\bold{0}$ at initial times.
Integrating out $\tilde{A}$ results in 
\begin{align}
\label{eq:zcf}
Z_\mathcal{C}^{\rm cl} = \int {[ {d\bar A} ]} \int {d\Pi _\bold{0}} \, {\rho _{W}}( {{\bar A}_{\bold{0}}},{\Pi _{\bold{0}}}) \, \delta[ \partial F[\bar{A}]-\bar{j} \, ] \ ,
\end{align}
where the argument of the delta function is the classical equation of motion 
\begin{align}
 \label{eq:eom}
 \partial_\mu \bar{F}^{\mu\nu}_n=\frac{g}{2}\operatorname{tr}\left\{\bracket{\left[\bar{\psi}_n,\psi_n\right]}_{\bar{A}} \gamma^\nu\right\} \, .
\end{align}
This equation is subjected to the initial conditions as specified by the Wigner function $\rho_W$. 
Observables are calculated as ensemble averages by numerically solving the classical field equations and sampling over the initial conditions according to
\begin{align}
\braket{O[\bar{A}]} = \int {[ {d\bar A} ]} \int {d\Pi _{\bold{0}}}  {\rho _{W}}({{\bar A}_{\bold{0}}},{\Pi _{\bold{0}}}) O[\bar{A}] \, \delta[ \partial F[\bar{A}]-\bar{j} ] \ .
\end{align}

\subsection{SU(N) gauge theory}
\label{sec:sun}
The derivation of the classical-statistical approximation for QCD or SU(N) gauge theory coupled to fermions proceeds along the same lines as for the U(1) case. Accordingly, we do not repeat the derivation but rather indicate where differences appear.

The fermions in the fundamental representation of the SU(N) gauge group carry the color index $j\in\{1,...,N\}$.
For the following discussion the link variables $U_{n,\mu}$ are expressed in terms of the gauge fields $A_{\mu,n}=t^a A^a_{\mu,n}$, where the $t^a$ are the generators of the SU(N) gauge group:
\begin{align}
 U_{\mu,n} = \exp \left( i g a_{\mu} A_{\mu,n} \right) \,.
\end{align}
For the non-Abelian group, the explicit parametrization of the Haar measure in terms of gauge fields introduces non-trivial contributions to the functional integral \cite{Kawai:1980ja}:
\begin{align} 
 \label{eq:GenFuncSU3}
 Z_\mathcal{C} = \int {\left[ {dA} \right]} \int {\left[ {d\bar \psi \, d\psi } \right]}\,\rho(t_0) \exp\left( i{S_G} + i{S_F} - {S_M}\right) \, , 
\end{align} 
where apart from the standard gauge and fermion actions $S_G + S_F$ also
\begin{align} \label{eq:HaarMeasureA}
 S_M = -\frac{1}{2} \sum \limits_{\substack{n\in\Lambda\\ \mu}} \mathrm{tr} \log \left[ 1 + N(A_{\mu,n})\right] \ , 
\end{align}
appears. Here the trace is taken over the adjoint gauge indices and
\begin{align} \label{eq:HaarMeasureB}
 N (A_{\mu,n}) = 2 \sum \limits_{l=1}^{\infty} \frac{(-1)^l}{(2l+2)!}(ga_{\mu}A_{\mu,n})^{2l} \, .
\end{align}

In order to derive the classical-statistical approximation of the quantum theory, we then proceed as in the Abelian case by expanding the exponent appearing in equation \eqref{eq:GenFuncSU3} in powers of the quantum field $\tilde{A}$. 
First we note that the contribution from $S_M$ that is linear in $\tilde{A}$ vanishes, as is shown in Appendix~\ref{app:B}, and the remaining zeroth order term can be identified with the functional integral measure of the classical field as $\int [d\bar{A}] \exp(-S_M[\bar{A}])$. The gauge action $S_G$ can be written as
\begin{align} 
 \label{eq:SU3GaugeActionOrder2}
 {S_G}[A] =- \frac{1}{2} \sum_{n \in \Lambda} a_{0}a_S^3\operatorname{tr}[F_{\mu \nu,n}[A] F_n^{\mu\nu}[A]] \ , 
\end{align}
up to higher orders in the lattice spacing, where the field strength is given by ${F}_{\mu\nu,n}=t^a{F}_{\mu\nu,n}^{a}$ with
\begin{align}
 {F}^{a}_{\mu\nu,n}[A] = \partial_{\mu} A^{a}_{\nu,n} - \partial_{\nu} A^{a}_{\mu,n} - g f^{abc} A^{b}_{\mu,n}  A^{c}_{\nu,n}   
\end{align}
and the trace is performed with respect to the adjoint gauge indices. Here $f^{abc}$ denote the structure constants of the SU(N) gauge group.
Again, we may rewrite the theory in terms of classical fields $\bar{A}$ and quantum fields $\tilde{A}$.
This results in a free part $S_1$ of the gauge action, an interacting part $S_2$ that is linear in $\tilde{A}$, and an interacting part $S_3$ non-linear in $\tilde{A}$:
\begin{align}
 \label{eq:nonAbelian}
 S_G[\bar{A}, \tilde{A}] = S_1 + S_2 + S_3 \, .
\end{align}
These different contributions are
\begin{widetext}
\begin{subequations} \label{eq:Verticesa}
\begin{align} 
S_1 =& \sum\limits_{n_1}{{\tilde A}^{a}_{\nu,n_1}}\partial_\mu\left[{\partial ^\mu}{\bar A_{n_1} }^{\nu,a} - {\partial ^\nu }{\bar A_{n_1} }^{\mu,a}\right] \, , \\
S_2 =& \frac{1}{2} \sum\limits_{n_1n_2n_3} V^{(3) \,abc}_{\mu\nu\rho} (n_1,n_2,n_3) \tilde{A}_{n_1}^{\mu,a}\bar{A}^{\nu,b}_{n_2}\bar{A}^{\rho,c}_{n_3} + 
       \frac{1}{6} \sum\limits_{n_1n_2n_3n_4} V^{(4) \,abcd}_{\mu\nu\rho\sigma} (n_1,n_2,n_3,n_4) \tilde{A}^{\mu,a}_{n_1}\bar{A}^{\nu,b}_{n_2}\bar{A}^{\rho,c}_{n_3}\bar{A}^{\sigma,d}_{n_4}\ , \label{eq:ClasQuantAction2} \\ 
S_3 =& \frac{1}{6} \sum\limits_{n_1n_2n_3} V^{(3) \,abc}_{\mu\nu\rho} (n_1,n_2,n_3) \tilde{A}^{\mu,a}_{n_1}\tilde{A}^{\nu,b}_{n_2}\tilde{A}^{\rho,c}_{n_3} + 
       \frac{1}{8}\sum\limits_{n_1n_2n_3n_4}V^{(4) \,abcd}_{\mu\nu\rho\sigma} (n_1,n_2,n_3,n_4) \tilde{A}^{\mu,a}_{n_1}\tilde{A}^{\nu,b}_{n_2} \tilde{A}^{\rho,c}_{n_3}\bar{A}^{\sigma,d}_{n_4} \ , \label{eq:ClasQuantAction3} 
\end{align}
\end{subequations}
\end{widetext}
where we suppressed the lattice spacings in the notation for simplicity.
Here $V^{(3)}$ and $V^{(4)}$ are the symmetrized three and four vertices of the gauge theory in terms of $\bar{A}$ and $\tilde{A}$ as given in Appendix~\ref{app:C}.

In order to obtain the classical-statistical approximation of the partition function, the contribution $S_3$ including the non-linear terms in the quantum field $\tilde{A}$ is neglected.
The fermionic contribution to the action is again obtained by integrating out the fermions, resulting in $\operatorname{Tr}\log\Delta_{\mathcal{C}}^{-1}[A]$ which is then expanded in the quantum field $\tilde{A}$.
The linear term in $\tilde{A}$ is proportional to the fermion current, which is now given by
\begin{align}
 \bar{j}^{a,\nu}(x) = \frac{g}{2}\mathrm{tr} \left\{ \bracket{\left[\bar{\psi}_n,\psi_n\right]}_{\bar{A}} \gamma^{\nu} t^{a} \right\} \ ,
\end{align}
where the trace is over Dirac and fundamental gauge indices.
The remainder of the derivation follows the U(1) case, such that the partition function for SU(N) gauge theory coupled to fermions in the classical-statistical approximation is again an expression equivalent to \eqref{eq:zcf}.
The equation of motion is now given by the classical Yang-Mills equation
\begin{equation}
 \partial_\mu \bar{F}^{\mu\nu,a}_n+gf^{abc}A^{b}_{\mu,n}\bar{F}^{\mu\nu,c}_{n}=\frac{g}{2}\operatorname{tr}\left\{\bracket{\left[\bar{\psi}_n,\psi_n\right]}_{\bar{A}} \gamma^\nu t^a\right\}\, .
\end{equation}

\section{Diagrammatics}
\label{sec:diag}

In this section we interpret the classical-statistical approximation of the underlying U(1) and SU(N) gauge theories coupled to fermions using a diagrammatic analysis. 

\begin{figure*}
 \includegraphics{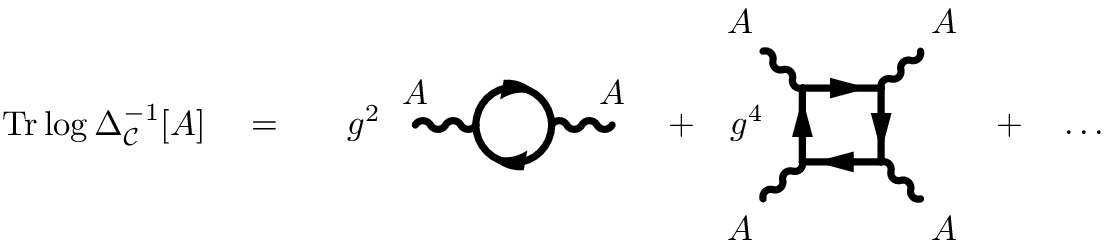}
 \caption{\label{fig:FullTrLog} Diagrammatic representation of $\mathrm{Tr} \log \Delta_{\mathcal{C}}^{-1}[A]$ as given in \eqref{eq:FermionTrLog}, with numerical prefactors being omitted. The fermion lines here denote free propagators $\Delta_{\mathcal{C}}[0]$.}
\end{figure*}

\subsection{Classical and quantum vertices} 

We start from \eqref{eq:IntegratingOutFermions} for the U(1) gauge theory with fermions integrated out and consider the expansion 
\begin{align} 
 \label{eq:FermionTrLog}
 \operatorname{Tr} \log &\Delta_{\mathcal{C}}^{-1}[A] =  
-\sum_{m=1}^{\infty}\frac{(ig)^{2m}}{2m}\mathrm{Tr}(\Delta_{\mathcal{C}}[0] \slashed{A})^{2m} \, , 
\end{align}
where we disregard constant and tadpole terms and note that only diagrams with an even number of photon lines contribute~\cite{Furry:1937zz}. 
The expansion is diagrammatically represented in Fig.~\ref{fig:FullTrLog}, showing that the $m$-th term in the sum corresponds to one fermionic loop with $2m$ photon lines attached. Expressing each term in the expansion (\ref{eq:FermionTrLog}) in terms of the classical field $\bar{A}$ and the quantum field $\tilde{A}$, the photon lines in Fig.~\ref{fig:FullTrLog} are then replaced by $\bar{A}$ or $\tilde{A}$, respectively.  

\begin{figure}[b]
\includegraphics{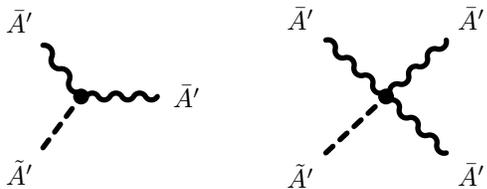}
\caption{\label{fig:classical} The classical vertices $S_2$ of the SU(N) gauge theory, which are independent of the coupling $g$ in terms of the rescaled fields $\bar{A}'$ and $\tilde{A}'$.}
\end{figure}

\begin{figure}[b]
\includegraphics{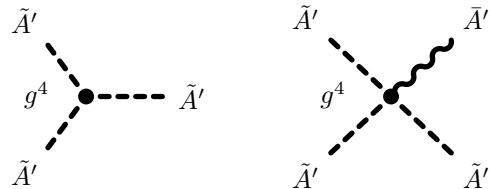}
\caption{\label{fig:quantum} The quantum vertices $S_3$ of the non-Abelian gauge theory, which are of order $g^4$ in terms of the rescaled fields $\bar{A}'$ and $\tilde{A}'$.}
\end{figure}

To proceed, we rescale these fields according to
\begin{subequations}
\begin{align} 
\bar{A} &= g^{-1} \bar{A'} \, , \\ 
\tilde{A} &= g \tilde{A'} \, . 
\end{align}
\label{eq:rescaling2}
\end{subequations}
The U(1) gauge action \eqref{eq:TildeAction} is invariant under this rescaling as it contains exactly one classical and one quantum field. 
The same is also true for the free part $S_1$ in the SU(N) gauge action \eqref{eq:Verticesa}.
Accordingly, these contributions do not depend on the coupling $g$.
In the non-Abelian gauge theory, however, we also have to consider the self-interactions as contained in $S_2$ and $S_3$. 
In fact, the contribution $S_2$ becomes also independent of $g$ after introducing the rescaled fields as is displayed in~Fig.~\ref{fig:classical}. This coupling independent part constitutes the classical vertex.
In contrast, the contribution $S_3$ corresponding to the quantum vertices becomes proportional to $g^4$ in the rescaled fields, cf.~Fig.~\ref{fig:quantum}.
In the classical-statistical approximation of pure gauge theories one disregards $S_3$ since these terms are non-linear in $\tilde{A}$. 
As a consequence, after the rescaling (\ref{eq:rescaling2}) the explicit coupling dependence drops out from the classical gauge dynamics and the coupling only enters via the initial conditions in this case.

\begin{figure*} 
\includegraphics{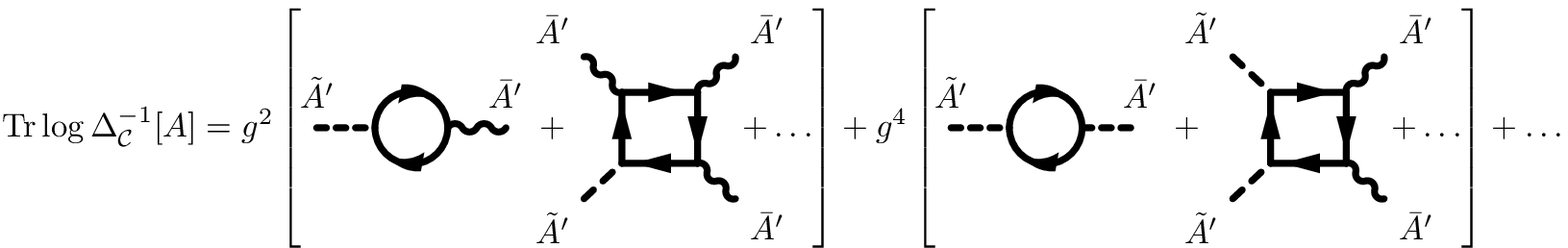}
\caption{\label{fig:current} Diagrammatic representation of Fig.~\ref{fig:FullTrLog} in terms of the rescaled field $\bar{A}'$ and $\tilde{A}'$. 
The term proportional to $g^2$ corresponds to the coupling of $\tilde{A}'$ to the fermion current \eqref{eq:currentcoupling}.
All higher order terms $\mathcal{O}(g^4)$ are neglected in the classical-statistical approximation.}
\end{figure*}

Taking into account the fermions, one observes that the coupling cannot be scaled out due to the intrinsic quantum nature of the fermions. 
To this end, we consider the coupling expansion of $\operatorname{Tr}\log \Delta_{\mathcal{C}}^{-1}[A]$ \eqref{eq:expansion} using  rescaled fields:
\begin{align}
 \label{eq:diagramatic}
  \operatorname{Tr} \log \,&\Delta_{\mathcal{C}}^{-1}[A] =\operatorname{Tr} \log \Delta_{\mathcal{C}}^{-1}[g^{-1}\bar{A}'] \nonumber\\
 & + g^2 \frac{i}{2} \mathrm{Tr}\{ \Delta_{\mathcal{C}}[g^{-1}\bar{A}'] \operatorname{sgn}_\mathcal{C} \slashed {\tilde {A}'}\} + \mathcal{O}(g^4) + \dots
\end{align}
One observes that the expansion in the coupling can also be understood as an expansion in the rescaled quantum field $\tilde{A}'$:
The first term $\operatorname{Tr} \log \Delta_{\mathcal{C}}^{-1}[g^{-1}\bar{A}']$ vanishes according to \eqref{eq:normalization}. 
The subsequent term proportional to $g^2$, on the other hand, contains all diagrams with only classical fields $\bar{A}'$ except for one quantum field $\tilde{A}'$.
In fact, these contributions sum up to the coupling of the quantum field $\tilde{A}'$ to the fermionic current \eqref{eq:currentcoupling}. The terms proportional to $g^4$ contain two quantum fields $\tilde{A}'$, the terms proportional to $g^6$ contain three quantum fields $\tilde{A}'$ and so on. The diagrammatic representation is shown in Fig.~\ref{fig:current}.
It has to be emphasized that already the contribution proportional to $g^2$ in the rescaled fields actually involves an infinite number of loops. 

In the classical-statistical approximation all terms in \eqref{eq:diagramatic} which contain more than one quantum field $\tilde{A}'$ are neglected. In return, this also means that the classical-statistical approximation is exact to order $g^2$ in the coupling whereas higher contributions $\mathcal{O}(g^4)$ are neglected. 
Consequently, the range of validity of the classical-statistical approximation is clearly restricted to small couplings 
$g\ll1$. In view of the restriction to weak couplings, it is important to note that a theory can be strongly correlated even for small couplings in the presence of large fields or occupation numbers. 
The classical-statistical approximation is well suited to discuss the corresponding nonperturbative physics in QED, or also QCD out of equilibrium at sufficiently small gauge coupling. The conditions for the validity of this description are discussed next.   

\subsection{Classicality condition}

The small couplings $g\ll1$, which are required for the validity of the classical-statistical approximation, involve large classical fields $\bar{A} \sim \mathcal{O}(1/g)$ according to (\ref{eq:rescaling2}).
For the initial-time problems considered in this work, these are implemented by the initial conditions for the subsequent nonequilibrium time evolution. As is explained in more detail in
Appendix~\ref{app:gauge_correlators}, we will consider (Coulomb gauge) Gaussian initial conditions for the discrete Fourier transformed gauge fields
\begin{align}
 \bar{A}_{i,(0,\mathbf{n})}\equiv \mathcal{A}_{i}+\frac{1}{V}\sum_{\mathbf{q}\in\tilde{\Lambda}^*}e^{i\mathbf{q}\cdot\mathbf{x}_n}  \bar{A}_{i,\mathbf{q}} \ ,
\end{align}
which may be characterized in terms of coherent fields 
\begin{subequations}
\begin{align}
 \langle \bar{A}_{i,(0,\mathbf{n})} \rangle &= \mathcal{A}_{i} \ , 
\end{align}
\end{subequations}
where we used $\langle  \bar{A}_{i,\mathbf{q}} \rangle=0$, as well as the connected two-point correlation function
\begin{subequations}
\begin{align} \left\langle  \bar{A}_{i,\mathbf{q}}  \bar{A}_{j,\mathbf{q}} \right\rangle & = \frac{V}{\omega_{\mathbf{q}}}\left(\frac{1}{2}+n_\mathbf{q}\right)\mathcal{P}_{ij} \ ,
\end{align}
\end{subequations}
and first-order time derivatives. Here $n_\mathbf{q}$ describes the initial occupation number with frequency $\omega_{\mathbf{q}}$ and $\mathcal{P}_{ij}$ denotes the transverse projector.
Vacuum initial conditions, i.e.\ quantum fluctuations around a coherent field, are determined by $n_{\mathbf{q}}=0$. 
The validity of the classical-statistical description in this case can be achieved by large initial coherent fields 
\begin{equation}
\mathcal{A}_{i} \sim \mathcal{O}(1/g) \, . \label{eq:ClasCriterion}
\end{equation}
We note that the relevant initial condition for Schwinger pair production, which will be considered in Sec.~\ref{sec:rtsim}, is expressed by a large first-order time derivative of the coherent field.
Alternatively, for small coherent fields one can consider large initial occupation numbers 
\begin{equation}
1 \ll n_\mathbf{q} \lesssim \mathcal{O}(1/g^2) \, ,
\end{equation}
for characteristic momenta $\mathbf{q}$ to achieve an accurate description of the underlying quantum dynamics.  
In general, during the time evolution large coherent fields or high occupation numbers can decrease such that time-dependent field amplitudes and occupancies have to be considered to monitor the range of validity of the classical-statistical simulations in time. 
Typically, the evolution has to be stopped once the characteristic time-dependent occupation numbers become of order one.  

In particular the late-time approach to thermal equilibrium, where the characteristic occupancies are of order one for typical momenta given by the temperature of the system, is beyond the range of validity of this approach. 
In this context it is also clear that a classical-statistical approximation suffers from Rayleigh-Jeans divergencies in the absence of an ultraviolet cutoff.
We emphasize that this represents no restriction for the use of the classical-statistical approximation to extract accurate results at sufficiently early times as long as the classicality conditions are respected: 
For given finite ultraviolet cutoff $\Lambda$ of the lattice regularized theory, the coupling has to be small enough such that the gauge field dynamics is dominated by large coherent fields or sufficiently highly occupied modes with typical momenta $|{\mathbf q}| \ll \Lambda$. 
In this case all results are insensitive to finite changes in the ultraviolet cutoff scale. 
This insensitivity can be used to verify the applicability of the classical-statistical approach, which has been described in detail in the context of scalar field theory~\cite{Berges:2013lsa}.        

We emphasize that the classicality condition only restricts the bosonic sector, where the analysis is in complete analogy to previous scalar field theory studies~\cite{Aarts:2001yn,Berges:2007ym}: More generally, the classicality condition is met whenever anti-commutator expectation values, such as $|\langle \bar{A}\bar{A} \rangle|$, for typical bosonic field modes are much larger than the corresponding commutators, such as $|\langle \tilde{A}\bar{A} - \bar{A} \tilde{A}\rangle|$. The latter involves one factor of the quantum field $\tilde{A}$ more than the corresponding anti-commutator. 
In principle, this allows one to discuss the classicality condition without explicit reference to the notion of occupancies.
However, the latter is very convenient and often not problematic in practice since the termination criterion typically checks for occupancies of order one at rather high momenta, where gauge fixed quantities in perturbation theory are often suitable. 
For an introductory review see also Ref.~\cite{Berges:2004yj}.

\section{Real-time simulation of quantum electrodynamics}
\label{sec:rtsim}

After deriving the classical-statistical approximation for gauge theories coupled to fermions, we now turn to the question of how to solve them on a computer. This amounts to solving an initial value problem numerically on a space-time lattice.
As indicated at the beginning of Sec.~\ref{sec:csftU1}, these simulations will be performed directly in terms of the link variables $U_{\mu,n}$ and not in terms of the gauge fields $A_{\mu,n}$.
Here we will restrict ourselves to U(1) gauge theory coupled to fermions, however, the generalization to SU(N) gauge theory follows along the very same lines.

\subsection{Lattice action}

In order to put the gauge fields on the lattice, we consider again the Wilsonian action on a real-time lattice \eqref{eq:gaugeaction}, which for the case of U(1) gauge symmetry can be written as
\begin{align}
 S_G\left[U\right] &= \frac{1}{g^2}\sum\limits_{n \in \Lambda^+}\sum_{i}{\frac{a^4}{a_0^2 a_i^2}}{\operatorname{Re}\mathrm{Tr}\left( {1 - {U_{0i,n}}} \right)}  \notag \\
 &-\frac{1}{g^2}\sum\limits_{n \in \Lambda^+}\sum_{\substack{i,j\\ i<j}}{\frac{a^4}{a_i^2a_j^2}}{\operatorname{Re}\mathrm{Tr}\left( {1 - {U_{ij,n}}} \right)}  \, ,
\end{align}
with the notation $a^4\equiv a_0a_1a_2a_3$. Here we allow for anisotropic lattices such that we distinguish between the different $a_i$ with $i\in\{1,2,3\}$. Moreover, we consider only the forward branch of the Schwinger-Keldysh contour according to the results of the derivation given in Sec.~\ref{sec:csft}. The electric and the magnetic field are then defined in terms of the temporal and spatial plaquettes, respectively:
\begin{subequations}
\begin{align}
 E_{i,n}&=\frac{1}{ga_0a_i}\operatorname{Im}{U_{0i,n}} \, , \\
 B_{i,n}&=-\frac{1}{2ga_ja_k}\epsilon_{ijk}\operatorname{Im}{U_{ij,n}} \ .
\end{align}
\end{subequations}

In the fermion sector, we employ the central derivative discretization as outlined in \eqref{eq:fermionaction}. 
In order to resolve the fermion doubling problem, which naturally arises in a lattice formulation of fermions \cite{Nielsen:1981}, several different approaches have been suggested \cite{Wilson:1974sk, Kogut:1975, Kaplan:1992bt, Frezzotti:2000nk}.
Here we employ Wilson fermions, which are most convenient in a theory without chiral symmetry, in order to treat the spurious doubler modes:
\begin{widetext}
 \begin{align}
 S_F[\psi,\bar{\psi},U] = 
 a^4\sum_{n\in\Lambda^+}{\bar{\psi}_n\left[i\gamma^\mu\frac{U_{\mu,n}\psi_{n+\hat{\mu}}-U_{-\mu,n}\psi_{n-\hat{\mu}}}{2a_\mu}-m\psi_n+\sum_{i}\frac{U_{i,n}\psi_{n+\hat{\imath}}-2\psi_n+U_{-i,n}\psi_{n-\hat{\imath}}}{2a_i}\right]} \ . 
 \end{align}
\end{widetext}
The last contribution corresponds to a second derivative term which vanishes in the naive continuum limit $a_i\to0$.
This Wilson term makes sure that the spatial doubler modes are suppressed, i.e.~it ensures that only low-momentum excitations show a low-energy dispersion relation.
We do not include a Wilson term for the temporal doubler modes as they are naturally suppressed for suitable initial conditions and if the temporal lattice spacing is taken to be much smaller than the spatial ones, $a_0\ll a_i$ \cite{Aarts:1998td,Borsanyi:2008eu,Mou:2013kca}.
Moreover, to simplify simulations afterwards, we use the gauge freedom and employ $U_{0,n}=1$, which is the lattice equivalent of the temporal-axial gauge condition $A_{0,n}=0$. 
\newpage
\subsection{Equations of motion}

Given the lattice action $S_G[U] + S_F[\psi,\bar{\psi},U]$, we can derive the lattice equations of motion by variation with respect to the dynamical degrees of freedom.
The equation of motion for the fermionic fields in temporal-axial gauge is then given by
\begin{widetext}
\begin{align}
 \label{eq:LatEomFerm}
 \psi_{n+\hat{0}}=\psi_{n-\hat{0}}-2ia_0\left(m+\sum_{i}\frac{1}{a_i}\right)\gamma^0\psi_n + \sum_{i}\frac{a_0}{a_i}\left[\left(i+\gamma^i\right)\gamma^0U_{i,n}\psi_{n+\hat{\imath}}+\left(i-\gamma^i\right)\gamma^0U_{-i,n}\psi_{n-\hat{\imath}}\right] \ .
\end{align}
\end{widetext}
In order to calculate the Dirac field $\psi_{n+\hat{0}}$ we have to know the link variables $U_{i,n}$ as well as the Dirac field at the two preceding time slices $\psi_{n}$ and $\psi_{n-\hat{0}}$.
The appearance of this leapfrog algorithm is an immediate consequence of the central derivative discretization of the Dirac action $S_F[\psi,\bar{\psi},U]$.
This also implies, that we have to choose two initial values for the Dirac field at two adjacent time slices $n_0=\{0,1\}$.
In an actual simulation, we choose $\psi_{(0,\mathbf{n})}$ according to a prescribed initial condition and then perform one step of a free field evolution of this initial condition to obtain a consistent value for $\psi_{(1,\mathbf{n})}$. 
In fact, it is exactly this choice of initial conditions which also keeps the temporal doubler mode unexcited \cite{Mou:2013kca}.

The classical equations of motion in the gauge sector \eqref{eq:eom} are formulated in terms of the Keldysh two-point function:
\begin{align}
 \Delta^{K}_{n,m}=\bracket{\left[\psi_n,\bar{\psi}_m\right]} \ .
\end{align}
It can be evaluated without further approximations using a mode function expansion as described in Appendix~\ref{app:mode_function} or employing a stochastic 'low-cost' fermion algorithm \cite{Borsanyi:2008eu,Berges:2013oba}.
In the former approach, the equations of motion \eqref{eq:LatEomFerm} are then regarded as equations of motion for the mode functions $\Phi^{u}_{\lambda,n,\mathbf{q}}$ and $\Phi^{v}_{\lambda,n,\mathbf{q}}$.

We calculate the Gauss law constraint in temporal axial gauge according to
\begin{align}
 \label{eq:LatEomGauss}
 \sum_{i}\frac{E_{i,n}-E_{i,n-\hat{\imath}}}{a_i}=-\frac{g}{2}\operatorname{Re}\operatorname{tr}\{\Delta^K_{n+\hat{0},n}\gamma^0\} \ ,
\end{align}
where the trace is over the Dirac indices.
This constraint equation has to be imposed on the initial field configuration in order to simulate in the physical subspace of the theory.
As a matter of fact, the constraint equation is conserved under the time evolution, i.e. a field configuration which fulfills Gauss law at initial times $n_0=0$ does also fulfill it at later times $n_0>0$.

The time evolution equation of the electric field, corresponding to Ampere's circuit law, is given by:
\begin{widetext}
\begin{align}
 \label{eq:LatEomEl}
  E_{i,n}=E_{i,n-\hat{0}}-\frac{a_0}{ga_i}\sum_{j\neq i}\frac{\operatorname{Im}[U_{ij,n}+U_{ji,n-\hat{\jmath}}]}{a_j^2} 
  +\frac{ga_0}{2}\operatorname{Re}\operatorname{tr}\{\Delta^{K}_{n+\hat{\imath}}(\gamma^i-i)U_{i,n}\} \ .
\end{align}
\end{widetext}

\subsection{Numerical algorithm}

The numerical algorithm can then be summarized in the following way:
\begin{itemize}
 \item[1.]{{\it Choose initial conditions} by specifying at times
 \begin{subequations}
 \begin{align}
  n_0=0:& \qquad \psi_{(0,\mathbf{n})} \ , \ E_{i,(0,\mathbf{n})} \ , \\ 
  n_1=0:& \qquad \psi_{(1,\mathbf{n})} \ , \ U_{i,(1,\mathbf{n})} \ .
 \end{align}
 \end{subequations}}
 \item[2.]{\it Solve the classical equations of motion}:
 \begin{itemize}
  \item[2a.]{{\it Dirac field evolution}: For a given $\psi_{n-0}$, $\psi_{n}$ and $U_{i,n}$ we obtain the Dirac field $\psi_{n+0}$ according to \eqref{eq:LatEomFerm}.}
  \item[2b.]{{\it Electric field evolution}: For a given $E_{i,n-\hat{0}}$, $\psi_{n}$ and $U_{i,n}$ we obtain the electric field $E_{i,n}$ according to \eqref{eq:LatEomEl}.}
  \item[2c.]{{\it Spatial link evolution}: We then explicitly calculate the temporal plaquette
  \begin{align}
   \quad\qquad U_{0i,n}=\sqrt{1-(ga_0a_iE_{i,n})^2}+iga_0a_iE_{i,n} \ , 
  \end{align}
  such that the link $U_{i,n+\hat{0}}$ is obtained as
  \begin{align}
   \qquad U_{i,n+\hat{0}}=U_{0i,n}U_{i,n} \ .
  \end{align}}
  \item[2d.]{Reiterate the steps $2a$ -- $2c$.}
 \end{itemize}
 \item[3.]{{\it{Perform a classical-statistical sampling}}: Reiterate the steps $1$ -- $2$ and average over the individual solutions to calculate observables.}
\end{itemize}

\subsection{Initial conditions \label{sec:Initial conditions}}

In order to solve the Cauchy problem, we have to specify initial conditions in both the fermion sector and the gauge sector.
In the following, we will assume that these two sectors decouple at initial times such that both of them can be considered as free.

In the fermion sector, we will restrict ourselves to the Dirac vacuum, which is characterized by the correlation functions:
\begin{subequations}
\begin{align}
 &\bracket{\psi_{(0,\mathbf{n})}}=\bracket{\bar{\psi}_{(0,\mathbf{n})}} = 0 \ . \\
 &\Delta^{K}_{(0,\mathbf{n}),(0,\mathbf{m})}=\frac{1}{V}\sum_{\mathbf{q}\in\tilde{\Lambda}}\frac{\bar{m}-\gamma^i\bar{p}_i}{\bar{\omega}}e^{i\mathbf{p}\cdot(\mathbf{x}_n-\mathbf{x}_m)} \ .
\end{align}
\end{subequations}
More details on these initial conditions can be found in Appendix~\ref{app:mode_function}.

In the gauge sector, we will assume a Gaussian density matrix, which is completely characterized by the one-point and two-point correlation functions of the dynamical degrees of freedom \cite{Cooper:1996ii}.
We have to choose a gauge condition in order to specify these initial correlations and consider temporal-axial gauge, where $A_{0,n}=0$ or equivalently $U_{0,n}=1$.
This gauge condition is incomplete since it leaves a residual gauge invariance under time-independent gauge transformations \cite{Leibbrandt:1987qv}.
We will use this residual gauge freedom at initial times $n_0=0$ to specify
\begin{align}
 \label{eq:gaugefreedom}
 \sum_{i}\frac{A_{i,(0,\mathbf{n})}-A_{i,(0,\mathbf{n}-\hat{\imath})}}{a_i}=0 \ .
\end{align}
Initial conditions, corresponding to a coherent field with vacuum fluctuations around it, are then expressed in terms of the one-point correlation functions:
\begin{subequations}
\begin{align}
 \bracket{A_{i,(0,\mathbf{n})}}&=\mathcal{A}_{i} \ , \\
 \bracket{E_{i,(0,\mathbf{n})}}&=\mathcal{E}_{i} \ ,
\end{align}
\end{subequations}
where $\mathcal{A}_{i}$ and $\mathcal{E}_{i}$ denote coherent fields, as well as the connected two-point correlation functions:
\begin{subequations}
\begin{align}
 \bracket{\{A_{i,(0,\mathbf{n})},A_{j,(0,\mathbf{m})}\}}_\text{c}&=\frac{1}{V}\sum_{\mathbf{q}\in\tilde{\Lambda}}\frac{1}{|\tilde{\mathbf{p}}|}\mathcal{P}_{ij}{e^{i\mathbf{p}\cdot(\mathbf{x}_n-\mathbf{x}_m)}} \\
 \bracket{\{E_{i,(0,\mathbf{n})},E_{j,(0,\mathbf{m})}\}}_\text{c}&=\frac{1}{V}\sum_{\mathbf{q}\in\tilde{\Lambda}}|\tilde{\mathbf{p}}|\mathcal{P}_{ij}{e^{i\mathbf{p}\cdot(\mathbf{x}_n-\mathbf{x}_m)}} \, 
 \end{align}
\end{subequations}
where $\mathcal{P}_{ij}$ denotes the transverse projector. 
More details about these initial conditions can be found in Appendix~\ref{app:gauge_correlators}.

Here, we initialize the vacuum modes only up to a finite momentum scale which is chosen to be well below the employed ultraviolet cutoff.
This ensures that the energy density of the vacuum modes is small such that they do not dominate the dynamics of the system.
We checked that the results are insensitive to this finite momentum scale, which has been chosen as $5m$ in the following.

\section{Fermion production simulations}
\label{sec:seff}
In this section, we consider as an application the production of electron-positron pairs by a large coherent field, the so-called Schwinger mechanism \cite{Sauter:1931zz,Heisenberg:1935qt,Schwinger:1951nm}.
To this end, we introduce the dimensionless field strength parameter
\begin{align}
 \epsilon_0=\frac{gE_0}{m^2} \ .
\end{align} 
For all subsequent numerical results we employ $g = 0.3$ as well as $\epsilon_0=3$. 

Regarding observables in the gauge sector, we are mainly interested in the one-point correlation function of the electric field.
In the fermion sector, we present results for the total fermion density $N(t)/V$, i.e.~the number of electrons per volume, as well as their normalized momentum distribution $n(p,t)$.

\subsection{Schwinger mechanism}

We first disregard the back-reaction of the fermion current on the gauge fields \eqref{eq:LatEomEl} as well as the classical-statistical sampling, i.e.~we do not choose non-trivial two-point correlation functions in the gauge sector.
We do this here to make contact to analytically known continuum results and to show that they can be reproduced with our real-time lattice simulations.
Starting with the vacuum initial conditions for the fermions, this approximation corresponds to only evolving the fermion equation of motion \eqref{eq:LatEomFerm} with a sudden switching-on of the electric field at $n_0=0$.

\begin{figure}[b]
\includegraphics[width=0.95\columnwidth]{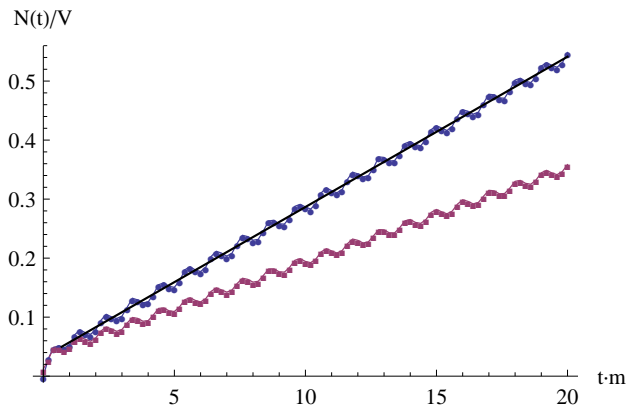}
 \caption{\label{fig:number_noback}Time evolution of the total fermion density $N(t)/V$ for $\epsilon_0=3$ with lattice parameters $a_0=0.002/m$, $a_3=0.05/m$, $N_{1,2}=12$, $N_3=40$.
Shown is a comparison of the continuum expression (solid line) with numerical results for $a_{1,2}=0.75/m$ (lower curve) and $a_{1,2}=0.25/m$ (upper curve).}
\end{figure}

In Fig.~\ref{fig:number_noback} we show the time evolution of the total fermion density, $N(t)/V$, for $\epsilon_0=3$ and two sets of different lattice spacings $a_i$.
We observe two different regimes: At early times of the order of $t_{\text{tr}}\sim1/m$ we observe a transient enhanced fermion production which can be attributed to the quench in the electric field.
At subsequent times, however, we observe a linear growth of the fermion density as expected from analytic continuum results, which are summarized in Appendix~\ref{app:schwinger}:
\begin{align}
 \frac{\dot{N}(t)}{V}=\frac{m^4 \epsilon_0^2}{4\pi^3}\exp\left(-\frac{\pi}{\epsilon_0}\right) \ .
\end{align}
We emphasize that for the derivation of this analytical result the initial time is sent to the remote past such that it cannot reproduce the transient regime. 
This analytic result is also shown in Fig.~\ref{fig:number_noback} for times after the transient regime.
We observe that our lattice result for $a_3=0.05/m$ and $a_{1,2}=0.25/m$ match the correct curve rather well whereas the result for $a_{1,2}=0.75/m$ still shows sizable deviations.
This demonstrates that the real-time lattice simulations can reproduce known results for small enough lattice spacings. Still, we have to be aware that we have deviations from the analytic results due to our numerical restriction to comparably small lattices.
Moreover, the oscillations in the numerical results around the analytical curve originates from the fact that we do not fully resolve the dynamics in momentum space for $N_3=40$ grid points.

\begin{figure}[t]
\includegraphics[width=0.95\columnwidth]{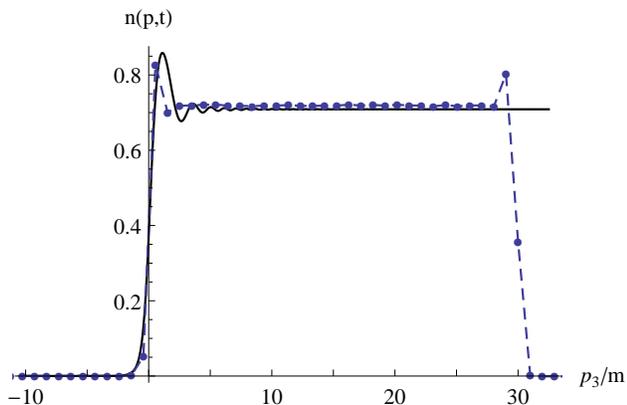}
 \caption{\label{fig:dist_noback} Comparison of the continuum expression (solid line) with the normalized momentum distribution $n(p,t)$ (dashed line) for $\epsilon_0=3$ at $p_{1,2}=0$ and $t=20/m$. 
 The parameters are $a_0=0.001/m$, $a_{1,2}=0.5/m$, $a_3=0.05/m$, $N_{1,2}=12$, $N_3=64$ such that $V=115.2/m^3$.}
\end{figure}

In Fig.~\ref{fig:dist_noback} we show the normalized momentum distribution $n(p,t)$ for $\epsilon_0=3$, $p_1=p_2=0$ at $t=20/m$ and compare it to the analytic continuum value $f(p)$, which is given in Appendix~\ref{app:schwinger}.
The common interpretation of $f(p)$ is such that electric field energy is transformed into virtual electron--positron pairs, showing up as the distinctive peak around kinetic momenta $p=0$.
For large enough field strengths, these charged excitations are separated over the Compton wavelength and become real electron--positron pairs. 
These real particles are then further accelerated by the electric field and, due to the neglect of back-reaction for the results in this section, gain momentum up to $p\to\infty$.

Obviously, we find good agreement between the numerical simulation and the analytic result, however, we observe a qualitatively different behavior for large momenta.
In fact, the analytic result assumes an electric field which has existed for all times such that all momenta up to $p\to\infty$ are occupied whereas the peak at large momenta in the numerical results is a consequence of the chosen initial conditions.
Because of the quench in the electric field, the enhanced production at early times $t_{\text{tr}}$ can be attributed to a single peak around $p=0$ which then propagates to higher and higher momenta during the time evolution.

\subsection{Back-reaction and plasma oscillations}

We now include the back-reaction of the fermion current on the gauge fields \eqref{eq:LatEomEl} as well as the classical-statistical sampling.
We find that  it suffices to take a very small number of field configurations as the physics is dominated by the large zero-mode of the electric field.
In fact, we performed five distinct runs which differed only marginally from each other upon calculating volume averaged quantities.

\begin{figure}[t]
 \includegraphics[width=0.95\columnwidth]{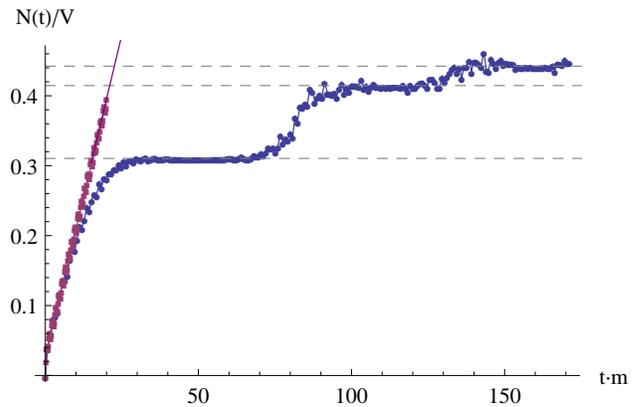}
 \caption{\label{fig:number_back}Time evolution of the total fermion density $N(t)/V$ for $\epsilon_0=3$ with lattice parameters $a_0=0.002/m$, $a_{1,2}=0.5/m$, $a_3=0.05/m$, $N_{1,2}=12$, $N_3=40$ such that $V=72/m^3$.
 For comparison, we include the straight line which shows the result without back-reaction.
 The horizontal dashed lines indicate the plateaus in the fermion density.}
\end{figure}

Fig.~\ref{fig:number_back} shows a comparison of the time evolution of the total fermion density $N(t)/V$ for $\epsilon_0=3$ in simulations with and without including the effect of the fermionic current.
We have already seen that the particle number grows linearly if we disregard the effect of the fermion current.
If we include this effect, however, the picture changes drastically as the fermion density assumes the shape of a staircase with decreasing step height.

\begin{figure}[b]
 \includegraphics[width=0.95\columnwidth]{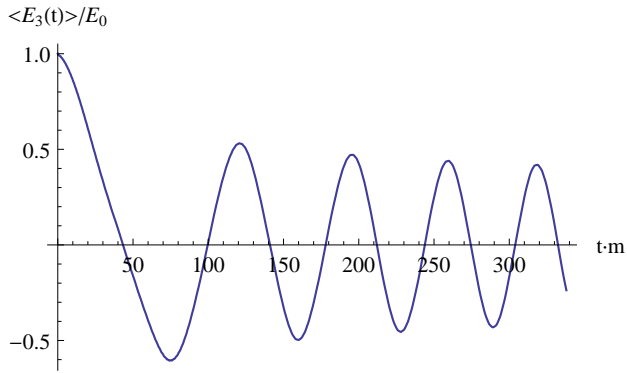}
 \caption{\label{fig:efld_back}Time evolution of the volume averaged electric field $\bracket{E_3(t)}$. The parameters are as in Fig.~\ref{fig:number_back}.}
\end{figure}

To understand this behavior, we also show the volume averaged electric field
\begin{align}
 \bracket{E_3(t)}=\frac{1}{V}\sum_{\mathbf{n}\in\Lambda} {E_{3,n}} 
\end{align}
for even longer times in Fig.~\ref{fig:efld_back}.
The expectation values of the other electric components $\bracket{E_{1,2}(t)}$ as well as the magnetic components $\bracket{B_i(t)}$ are equal to zero.
Starting from $t=0$, electron-positron pairs are created and accelerated such that a fermion current arises.
Accordingly, an electric field counteracting the initial electric field is formed.
As a consequence, the electric field eventually changes sign and grows until a first local minimum is achieved.
The electric field then increases again, changes sign, reaches a local maximum and so forth.
The occurrence of these plasma oscillations is in accordance with previous alternative investigations \cite{Hebenstreit:2013qxa,Kluger:1992gb,Bloch:1999eu}.

The fermion sector follows the oscillatory behavior of the electric field:
Particle production effectively terminates when the magnitude of the field strength becomes too small, corresponding to the approximate plateaus in $N(t)/V$.
However, at those instants of time at which the electric field reaches local extrema, electron-positron production sets in again.
Due to the fact that the envelope of the electric field decreases with time, $N(t)/V$ assumes the shape of a staircase with decreasing step height. 
We note that the oscillation frequency of the electric field increases with the number of produced fermions, in accordance with the expected parametric dependence. 

\begin{figure}[t]
\includegraphics[width=0.95\columnwidth]{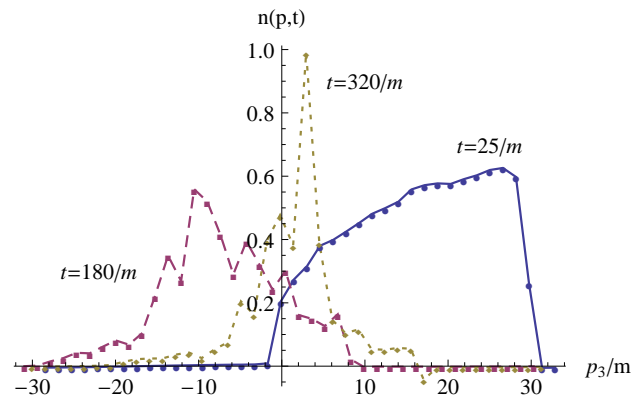}
 \caption{\label{fig:dist_back}Normalized momentum distribution $n(p,t)$ for $\epsilon_0=3$ at $p_{1,2}=0$. The parameters are as in Fig.~\ref{fig:number_back}.} 
\end{figure}

Finally, in Fig.~\ref{fig:dist_back} we show the normalized momentum distribution $n(p,t)$ at different times.
Due to the fact that the electric field changes sign again and again, the electrons and positrons are also accelerated back and forth in momentum space.
At later times, this results in a peaked distribution which oscillates around $p=0$ in accordance with the electric field.

\section{Conclusion}
\label{sec:conc}

In this work we investigated real-time lattice gauge theories coupled to fermions and showed how simulations can be performed in the classical-statistical regime, where the approach provides a nonperturbative description of the underlying quantum theory. 
We employed this method to study Schwinger pair production in quantum electrodynamics in three spatial dimensions. 
The comparison to analytic results for constant background field demonstrates the importance of back-reaction of the produced fermion pairs on the gauge fields at later times. 
On the other hand, at earlier times the comparison showed a very good agreement with the Schwinger formula pointing out the ability of the lattice simulation method to accurately describe the underlying pair creation process. 

The Dirac-Heisenberg-Wigner formalism provides an alternative approach to simulation pair production in inhomogeneous electromagnetic background fields \cite{BialynickiBirula:1991tx}.
It has to be emphasized, however, that actual simulations within this approach have been restricted to rather simple space- and time-dependent electric fields \cite{Hebenstreit:2011wk,Berenyi:2013eia}.
Moreover, the effect of back-reaction has been disregarded.
In this respect, real-time lattice gauge theory provides a unique tool to go beyond previous approximations and investigate the pair production process in more realistic space- and time-dependent electromagnetic fields.
The simulation of such field configurations is beyond the scope of this publication but will be addressed elsewhere.

Using a diagrammatic expansion we discussed the corrections coming from the inclusion of the fermion sector, which is genuinely quantum, and described the range of validity of the approach: 
The latter is given by the classicality condition for the gauge sector, which restricts the applications to the physics of large coherent fields or high typical occupancies. 
The same techniques can be applied to similar conditions for quark production in QCD with the important difference that gluon self-interactions are expected to play a dominant role in this case.
This investigation is deferred to a future publication, where also the role of large inhomogeneous fields or corresponding initial fluctuations will be considered.

\subsection*{Acknowledgments}

We thank D.~Gelfand, I.I. Rocca, S.~Schlichting and D.~Sexty for helpful  
discussions and collaborations on related work. F.~Hebenstreit is  
supported by the Alexander-von-Humboldt Foundation.

\appendix

\section{Fermionic current}
\label{app:A}
In order to rewrite the second term in \eqref{eq:expansion}, we introduce a continuum notation. 
The inverse propagator is then 
\begin{align}
 i\Delta^{-1}_\mathcal{C}[\bar{A}] \equiv \left(i \slashed{\partial}_x - g\slashed{\bar{A}}(x)-m \right)\delta_\mathcal{C}(x,y) \, ,
\end{align}
where higher terms in the lattice spacing are neglected and $\delta _{\mathcal{C}} \left(x,y\right)$ is the delta function on the time contour.
The subscript $\mathcal{C}$ indicates that the Schwinger-Keldysh contour is incorporated in this continuum notation. 
Accordingly, the propagator is determined by
\begin{align} 
 \label{eq:ContProp} 
 \left(i \slashed{\partial}_x - g \slashed{\bar{A}}(x)-m\right)\Delta _{\mathcal{C}}\left(x,y\right)=i \delta _{\mathcal{C}} \left(x,y\right) \ .
\end{align}
The connection between contour-ordered correlation functions and the propagator is given as
\begin{align}
 \Delta_{\mathcal{C}}\left(x ,y \right)=\left\langle T_\mathcal{C}\psi(x)\bar{\psi}(y)\right\rangle_{\bar{A}} \ ,
\end{align}
where $T_\mathcal{C}$ denotes time ordering along the Schwinger-Keldysh contour.
The notation $\langle\cdots\rangle_{\bar{A}}$ indicates that these correlation functions are determined by \eqref{eq:ContProp} for a given classical field $\bar{A}$.
The propagator can be written according to
\begin{align} 
 \Delta _{\mathcal{C}} \left(x,y\right) =\theta _{\mathcal{C}} \left(x_0,y_0\right)\Delta^{>} \left(x,y\right)+\theta _{\mathcal{C}} \left(y_0,x_0\right)\Delta^{<}\left(x,y\right) \, ,
\end{align} 
where $\theta_\mathcal{C}(x_0,y_0)$ is the generalization of the Heaviside function on the Schwinger-Keldysh contour \cite{Kobes:1985}.
The different components are given by
\begin{subequations}
\begin{align} 
 \Delta^{>} \left(x,y\right)&= \bracket{\psi (x)\bar{\psi}(y)}_{\bar{A}} \, ,\\ 
 \Delta^{<} \left(x,y\right)&=-\bracket{ \bar{\psi }\left(y\right)\psi \left(x\right)}_{\bar{A}} \, ,
\end{align} 
\end{subequations}
such that an explicit evaluation of the trace in \eqref{eq:expansion} gives
\begin{align} \label{eq:FirstOrderATilde}
 \operatorname{Tr} \{\Delta_{\mathcal{C}}[\bar{A}] \operatorname{sgn}_\mathcal{C} \slashed {\tilde {A}}\} = \operatorname{tr}\int_{t_0}^{t_F}\int\limits_{\mathbf{x}}[\Delta^{>}(x,x)+\Delta^{<}(x,x)]\slashed{\tilde{A}}(x) \, .
\end{align}
Here $\operatorname{tr}$ denotes the trace over Dirac indices and we employed the relation
\begin{align}
 {\theta_\mathcal{C}}\left( x_0,y_0 \right) + {\theta_\mathcal{C}}\left( y_0,x_0 \right) = 1 \, .
\end{align}
The Keldysh propagator of fermionic fields is defined according to
\begin{align}
 \Delta^K(x,y)\equiv\Delta^{>}(x,y)+\Delta^{<}(x,y)=\left\langle{\left[\psi(x),\bar{\psi}(y)\right]}\right\rangle_{\bar{A}} \, ,
\end{align}
and fulfills
\begin{align}
 \left(i \slashed{\partial}_x - g \slashed{\bar{A}}(x)-m\right)\Delta^K\left(x,y\right)=0 \ .
\end{align}
Accordingly, \eqref{eq:FirstOrderATilde} can be written as:
\begin{align} 
 \label{eq:current}
 &\operatorname{Tr}  \{\Delta_{\mathcal{C}}[\bar{A}] \operatorname{sgn}_\mathcal{C} \slashed{\tilde{A}}\} = 
  \operatorname{tr} \int_{t_0}^{t_F}\int\limits_{\mathbf{x}}\Delta^K(x,x)\, \slashed{\tilde{A}}(x) \, .
\end{align}
It has to be emphasized that the term on the right hand side of this equation is closely related to the current of fermionic particles coupled to the quantum field $\tilde{A}$ \cite{Cooper:1994hr},
\begin{align}
 \bar{j}^{\nu}(x) = - \frac{g}{2}\operatorname{tr}\left\{\Delta^K(x,x) \gamma^\nu\right\} \, ,
\end{align}
where the trace $\operatorname{tr}$ is taken over the Dirac indices.

\section{Haar measure}
\label{app:B}
We start from the explicit expression for the Haar measure of SU(N) gauge theories given by \eqref{eq:HaarMeasureA} and \eqref{eq:HaarMeasureB}.
Expanding $S_M$ in the quantum field $\tilde{A}$ gives,
\begin{align}
 S_M=&-\frac{1}{2} \sum_{\substack{n\in\Lambda\\ \mu  }} \mathrm{tr} \log [1+ N({A}_{\mu,n})] \nonumber \\
    =&-\frac{1}{2} \sum_{\substack{n\in\Lambda\\ \mu  }} \mathrm{tr} \log (1+ N({\bar{A}}_{\mu,n})) \nonumber \\
     &-\frac{1}{2} \sum_{\substack{n\in\Lambda\\ \mu  }} B^{a}({\bar{A}}_{\mu,n})\operatorname{sgn}_{\mathcal{C}} \tilde{A}^{a}_{\mu,n} +\dots\ ,
\end{align}
where we used $A=\bar{A}+\frac{1}{2}\operatorname{sgn}_\mathcal{C}\tilde{A}$.
Here, $B^{a}(\bar{A}_{\mu,n})$ is given by the first derivative of the logarithm with respect to $\tilde{A}$ evaluated at $\tilde{A} = 0$. Hence
it is a local function of the classical field ${\bar{A}}$. 
Since this function has the same value on the forward and backward branch whereas the $\operatorname{sgn}_\mathcal{C}$ changes its sign, the first order term vanishes.
Due to the fact that the linear terms in the quantum field determine the classical equations of motion, the Haar measure does not contribute to them. 

The remaining zeroth order term can be identified with the Haar measure of the classical field: 
\begin{align}
\int [d\bar{A}] \exp(-S_M[\bar{A}]) = \int [d\bar{U}] \, ,
\end{align}
which is in agreement with the definition of the expectation value of observables given in \eqref{eq:observables}. 

\section{Vertices of the SU(N) gauge theory}
\label{app:C}

For the sake of completeness, we state the continuum expressions for the symmetrized three vertex:
\begin{align}
&V^{(3) \,abc}_{\mu\nu\rho} (x_1,x_2,x_3) = \notag \\
&\hphantom{+\,\,\,} g f^{abc} \eta_{\mu \nu} (\delta_{x_2,x_3} \partial_\rho^{x_1} \delta_{x_1,x_2}- \delta_{x_1,x_3} \partial_\rho^{x_2} \delta_{x_2,x_1})\notag \\
&+g f^{abc} \eta_{\mu\rho}(\delta_{x_1,x_2} \partial_\nu^{x_3} \delta_{x_3,x_1} 
-\delta_{x_2,x_3} \partial_\nu^{x_1} \delta_{x_1,x_3}) \notag \\ 
&+g f^{abc} \eta_{\nu \rho}(\delta_{x_1,x_3} \partial_\mu^{x_2} \delta_{x_2,x_1} 
-\delta_{x_1,x_2} \partial_\mu^{x_3} \delta_{x_3,x_1} ) \ ,
\end{align}
as well as the the four vertex:
\begin{align}
&V^{(4) \,abc}_{\mu\nu\rho\sigma} (x_1,x_2,x_3,x_4) = \delta_{x,y} \delta_{x,w} \delta_{x,z}\notag \\
&\times\left[- \frac{1}{4} g^2 f^{abe} f^{cde} (\eta_{\rho\mu}
\eta_{\sigma\nu}-\eta_{\mu\sigma}\eta_{\nu\rho}) \right. \notag \\
&\hphantom{\times\left[\vphantom{\frac{1}{4}} \right.}- \frac{1}{4} g^2 f^{ace}
f^{bde} (\eta_{\mu\nu} \eta_{\sigma\rho}-\eta_{\mu\sigma}\eta_{\nu\rho}) \notag \\
&\hphantom{\times\left[\vphantom{\frac{1}{4}} \right.}\left.- \frac{1}{4} g^2 f^{ade}
f^{cbe} (\eta_{\mu\rho} \eta_{\sigma\nu}-\eta_{\mu\nu}\eta_{\rho\sigma}) \right]\ .
\end{align}
of SU(N) gauge theory.
Here, $\eta=\text{diag}(1,-1,-1,-1)$ denotes the Minkowski metric.
The corresponding lattice expressions for $V^{(3)}$ and $V^{(4)}$ can be found in \cite{Rothe:1992nt}. 

\section{Mode function expansion for fermions}
\label{app:mode_function}

In the mode function expansion, the Dirac field operator is expressed in terms of time-dependent mode functions $\Phi^{u}_{\lambda,n,\mathbf{q}}$, $\Phi^{v}_{\lambda,n,\mathbf{q}}$ and time-independent creation/annihilation operators $b_{\lambda,\mathbf{q}}$, $d^\dagger_{\lambda,\mathbf{q}}$:
\begin{align}
 \psi_n=\frac{1}{V}\sum_{\mathbf{q}\in\tilde{\Lambda}}\sum_{\lambda}\left[\Phi^{u}_{\lambda,n,\mathbf{q}}b_{\lambda,\mathbf{q}}+\Phi^{v}_{\lambda,n,\mathbf{q}}d^{\dagger}_{\lambda,\mathbf{q}}\right] \ ,
\end{align}
with the total volume $V=\prod_i N_ia_i$ and the spin index $\lambda\in\{1,2\}$. 
The conjugate lattice $\tilde{\Lambda}$ is defined as
\begin{align}
 \label{eq:conjglat}
 \tilde{\Lambda} = \left\{\mathbf{q} \left|\, q_i=\frac{N_ia_ip_i}{2\pi} \in \-\frac{N_i}{2}, \dots , \frac{N_i}{2}-1 \right\}\right. \ ,
\end{align}
where we assumed periodic boundary conditions in the spatial directions. 
The creation/annihilation operators obey
\begin{align}
 \big\{b_{\lambda,\mathbf{q}},b^\dagger_{\lambda',\mathbf{q}'}\big\}=\big\{d_{\lambda,\mathbf{q}},d^\dagger_{\lambda',\mathbf{q}'}\big\}=V\delta_{\lambda,\lambda'}\delta_{\mathbf{q},\mathbf{q}'} \ .
\end{align}
The fermion occupation numbers are determined by
\begin{subequations}
\begin{align}
 \big\langle b^\dagger_{\lambda,\mathbf{q}}b_{\lambda,\mathbf{q}}\big\rangle=Vn^{u}_{\lambda,\mathbf{q}} \ , \\
 \big\langle d^\dagger_{\lambda,\mathbf{q}}d_{\lambda,\mathbf{q}}\big\rangle=Vn^{v}_{\lambda,\mathbf{q}} \ ,
\end{align}
\end{subequations}
where we assume a decoupling of the fermion and the gauge sector at initial times $n_0=0$, such that
\begin{subequations}
\begin{align}
 \Phi^u_{\lambda,(0,\mathbf{n}),\mathbf{q}}&=u_{\lambda,\mathbf{q}}e^{i\mathbf{p}\cdot\mathbf{x}_n}\ , \\
 \Phi^v_{\lambda,(0,\mathbf{n}),\mathbf{q}}&=v_{\lambda,\mathbf{q}}e^{-i\mathbf{p}\cdot\mathbf{x}_n} \ ,
\end{align}
\end{subequations}
Here, $\mathbf{x}_n=(a_1n_1,a_2n_2,a_3n_3)$ and $\mathbf{p}=(p_1,p_2,p_3)$, with $p_i$ defined in \eqref{eq:conjglat}. 
Employing the Dirac representation of the $\gamma$-matrices
\begin{equation}
  \gamma^0=\begin{pmatrix}\mathbb{1}&0\\0&-\mathbb{1}\end{pmatrix} \qquad , \qquad \gamma^i=\begin{pmatrix}0&\sigma^i\\-\sigma^i&0\end{pmatrix} 
\end{equation}
an explicit representation of the free eigenspinors $u_{\lambda,\mathbf{q}}$ and $v_{\lambda,\mathbf{p}}$ is given by
\begin{subequations}
\begin{align}
  u_{1,\mathbf{q}}=\sqrt{\tfrac{\bar{\omega}+\bar{m}}{2\bar{\omega}}}\begin{pmatrix}1\\0\\\tfrac{\bar{p}_3}{\bar{\omega}+\bar{m}}\\\tfrac{\bar{p}_1+i\bar{p}_2}{\bar{\omega}+\bar{m}}\end{pmatrix}  &\ , \   u_{2,\mathbf{q}}=\sqrt{\tfrac{\bar{\omega}+\bar{m}}{2\bar{\omega}}}\begin{pmatrix}0\\1\\\tfrac{\bar{p}_1-i\bar{p}_2}{\bar{\omega}+\bar{m}}\\\tfrac{-\bar{p}_3}{\bar{\omega}+\bar{m}}\end{pmatrix} \\
  v_{1,\mathbf{q}}=\sqrt{\tfrac{\bar{\omega}+\bar{m}}{2\bar{\omega}}}\begin{pmatrix}\tfrac{\bar{p}_3}{\bar{\omega}+\bar{m}}\\\tfrac{\bar{p}_1+i\bar{p}_2}{\bar{\omega}+\bar{m}}\\1\\0\end{pmatrix}  &\ , \   v_{2,\mathbf{q}}=\sqrt{\tfrac{\bar{\omega}+\bar{m}}{2\bar{\omega}}}\begin{pmatrix}\tfrac{\bar{p}_1-i\bar{p}_2}{\bar{\omega}+\bar{m}}\\\tfrac{-\bar{p}_3}{\bar{\omega}+\bar{m}}\\0\\1\end{pmatrix}
\end{align}
\end{subequations}
with
\begin{subequations}
\label{eq:lat_mom}
\begin{align}
 \bar{p}_i&=\frac{1}{a_i}\sin\left(\frac{2\pi q_i}{N_i}\right) \ , \\
 \bar{m}&=m+\sum_{i}\frac{2}{a_i}\sin^2\left(\frac{\pi q_i}{N_i}\right) \ , \\
 \bar{\omega}&=\sqrt{\bar{m}^2+\bar{p}_1^2+\bar{p}_2^2+\bar{p}_3^2} \ .
\end{align}
\end{subequations}
At any later time $n_0,m_0>0$, the Keldysh propagator is calculated according to
\begin{align}
 \Delta^K_{n,m}=\frac{1}{V}\sum_{\mathbf{q}\in\tilde{\Lambda}}\sum_{\lambda}&{\left[\Phi^{u}_{\lambda,n,\mathbf{q}}\bar{\Phi}^{u}_{\lambda,m,\mathbf{q}}(1-2n^{u}_{\lambda,\mathbf{q}})\right.} \nonumber \\
									   &\left.-\,\Phi^{v}_{\lambda,n,\mathbf{q}}\bar{\Phi}^{v}_{\lambda,m,\mathbf{q}}(1-2n^{v}_{\lambda,\mathbf{q}})\right] \ ,
\end{align}
with the mode functions obeying the equation of motion \eqref{eq:LatEomFerm}
and $\bar{\Phi} = \Phi^{\dagger} \gamma^{0}$.
We note that vacuum initial conditions $n^{u}_{\lambda,\mathbf{q}}=n^{v}_{\lambda,\mathbf{q}}=0$ are specified by the following one-point correlation functions and the Keldysh propagator:
\begin{subequations}
\begin{align}
 &\bracket{\psi_{(0,\mathbf{n})}}=\bracket{\bar{\psi}_{(0,\mathbf{n})}}=0 \ , \\
 &\Delta^{K}_{(0,\mathbf{n}),(0,\mathbf{m})}=\frac{1}{V}\sum_{\mathbf{q}\in\tilde{\Lambda}}\frac{\bar{m}-\gamma^i\bar{p}_i}{\bar{\omega}}e^{i\mathbf{p}\cdot(\mathbf{x}_n-\mathbf{x}_m)} \ ,
\end{align}
\end{subequations}
where we used $\langle b_{\lambda,\mathbf{q}}\rangle=\langle d_{\lambda,\mathbf{q}}\rangle=0$.

\section{Mode function expansion for gauge fields}
\label{app:gauge_correlators}

In order to solve the Gauss law constraint for the Dirac vacuum \eqref{eq:LatEomGauss} and the residual gauge condition \eqref{eq:gaugefreedom}, we perform a discrete Fourier transformation
\begin{align}
 \label{eq:Fourier}
 E_{i,(0,\mathbf{n})}\equiv \mathcal{E}_{i}+\frac{1}{V}\sum_{\mathbf{q}\in{\tilde{\Lambda}}^*}e^{i\mathbf{p}\cdot\mathbf{x}_n}{E}_{i,\mathbf{q}} \ ,
\end{align}
with $ \tilde{\Lambda}^* = \tilde{\Lambda} \setminus \left\{ \mathbf{q} = \mathbf{0} \right\}$, and similarly for $A_{i,(0,\mathbf{n})}$. 
Here, $\mathcal{E}_{i}$ denotes the coherent field in the zero-momentum mode.
Accordingly, the transversality condition in conjugate space reads
\begin{align}
 \label{eq:transv}
 \sum_{i}\tilde{p}_i{E}_{i,\mathbf{q}}=0=\sum_{i}\tilde{p}_i{A}_{i,\mathbf{q}} \ ,
\end{align}
with
\begin{subequations}
\begin{align}
 \tilde{p}_i&=\frac{2}{a_i}e^{-i\pi q_i/N_i}\sin\left(\frac{\pi q_i}{N_i}\right) \ , \\
 |\tilde{\mathbf{p}}|&=\sqrt{\tilde{p}_1^2+\tilde{p}_2^2+\tilde{p}_3^2} \ . 
\end{align}
\end{subequations}
We solve \eqref{eq:transv} explicitly in terms of a mode function expansion:
\begin{subequations}
\begin{align}
 {A}_{i,\mathbf{q}}&=\frac{1}{\sqrt{2|\tilde{\mathbf{p}}|}}\sum_{\lambda}\big[a_{\lambda,\mathbf{q}}\epsilon_{i,\lambda,\mathbf{q}}+a_{\lambda,-\mathbf{q}}^\dagger\epsilon^*_{i,\lambda,-\mathbf{q}}\big] \ , \\ 
 {E}_{i,\mathbf{q}}&=i\sqrt{\frac{|\tilde{\mathbf{p}}|}{2}}\sum_{\lambda}\big[a_{\lambda,\mathbf{q}}\epsilon_{i,\lambda,\mathbf{q}}-a_{\lambda,-\mathbf{q}}^\dagger\epsilon^*_{i,\lambda,-\mathbf{q}}\big] \ , 
\end{align}
\end{subequations}
with polarization vectors $\bm{\epsilon}_{\lambda,\mathbf{q}}$ and polarization index $\lambda\in\{1,2\}$.
The creation/annihilation operators obey the non-trivial commutation relation
\begin{equation}
 \big[a_{\lambda,\mathbf{q}},a_{\lambda',\mathbf{q}'}^\dagger\big]=V\delta_{\lambda\lambda'}\delta_{\mathbf{q}\mathbf{q}'} \ .
\end{equation}
The gauge field occupation numbers are determined by
\begin{align}
 \big\langle a^\dagger_{\lambda,\mathbf{q}}a_{\lambda,\mathbf{q}}\big\rangle=Vn_{\lambda,\mathbf{q}} \ .
\end{align}
The transversality \eqref{eq:transv} condition is trivially fulfilled for polarization vectors obeying
\begin{subequations}
\begin{align}
 \tilde{\mathbf{p}}\cdot\bm{\epsilon}_{\lambda,\mathbf{q}}&=0 \ , \\
 \bm{\epsilon}_{\lambda,\mathbf{q}}^*\cdot\bm{\epsilon}_{\lambda',\mathbf{q}}&=\delta_{\lambda\lambda'} \ .
\end{align}
\end{subequations}
In fact, we may construct an explicit representation for the polarization vectors. For $q_1\neq0$ we use:
\begin{subequations}
\begin{align}
 \bm{\epsilon}_{1,\mathbf{q}}&=\frac{1}{\sqrt{|\tilde{p}_1|^2+|\tilde{p}_2|^2}}\begin{pmatrix}-\tilde{p}_2\\\tilde{p}_1\\0\end{pmatrix} \ , \\
 \bm{\epsilon}_{2,\mathbf{q}}&=\frac{1}{|\tilde{\mathbf{p}}|\sqrt{|\tilde{p}_1|^2+|\tilde{p}_2|^2}}\begin{pmatrix}\tilde{p}_1^*\tilde{p}_3\\\tilde{p}_2^*\tilde{p}_3\\-|\tilde{p}_1|^2-|\tilde{p}_2|^2\end{pmatrix} \ ,
\end{align}
\end{subequations}
whereas for $q_1=0$ we employ:
\begin{align}
 \bm{\epsilon}_{1,\mathbf{q}}&=\frac{1}{\sqrt{|\tilde{p}_2|^2+|\tilde{p}_3|^2}}\begin{pmatrix}0\\\tilde{p}_3\\-\tilde{p}_2\end{pmatrix} \ \ , \ \ \bm{\epsilon}_{2,\mathbf{q}}=\begin{pmatrix}1\\0\\0\end{pmatrix} \ .
\end{align}
In this representation, the polarization vectors fulfill
\begin{subequations}
\begin{align}
 \bm{\epsilon}_{1,-\mathbf{q}}^*&=-\bm{\epsilon}_{1,\mathbf{q}} \ , \\
 \bm{\epsilon}_{2,-\mathbf{q}}^*&=\bm{\epsilon}_{2,\mathbf{q}} \ ,
\end{align}
\end{subequations}
such that the transverse projector $\mathcal{P}$ is given by
\begin{equation}
\label{eq:transvp} \mathcal{P}_{ij}=\sum_{\lambda}\epsilon_{i,\lambda,\mathbf{q}}\epsilon^*_{j,\lambda,\mathbf{q}}=\delta_{ij}-\frac{\tilde{p}_i\tilde{p}_j^*}{|\tilde{\mathbf{p}}|^2} \ .
\end{equation}
A Gaussian initial state is then specified in terms of the one-point correlation functions, corresponding to coherent background fields:
\begin{subequations}
\begin{align}
 \langle A_{i,(0,\mathbf{n})} \rangle &= \mathcal{A}_{i} \ , \\
 \langle E_{i,(0,\mathbf{n})} \rangle &= \mathcal{E}_{i} \ ,
\end{align}
\end{subequations}
where we used $\langle a_{\lambda,\mathbf{q}}\rangle=0$, as well as the connected two-point correlation functions:
\begin{subequations}
\begin{align}
 &\frac{1}{2}\bracket{\{A_{i,(0,\mathbf{n})},A_{j,(0,\mathbf{m})}\}}-\langle A_{i,(0,\mathbf{n})} \rangle\langle A_{j,(0,\mathbf{m})} \rangle \ , \\ 
 &\frac{1}{2}\bracket{\{A_{i,(0,\mathbf{n})},E_{j,(0,\mathbf{m})}\}}-\langle A_{i,(0,\mathbf{n})} \rangle\langle E_{j,(0,\mathbf{m})} \rangle \ , \\ 
 &\frac{1}{2}\bracket{\{E_{i,(0,\mathbf{n})},E_{j,(0,\mathbf{m})}\}}-\langle E_{i,(0,\mathbf{n})} \rangle\langle E_{j,(0,\mathbf{m})} \rangle \ . 
\end{align}
\end{subequations}
Employing the discrete Fourier decomposition \eqref{eq:Fourier} and assuming $n_{1,\mathbf{q}}=n_{2,\mathbf{q}}\equiv n_{\mathbf{q}}$, we obtain
\begin{subequations}
\begin{align}
 \frac{1}{2}\big\langle\{{A}_{i,\mathbf{q}},{A}_{j,\mathbf{q}}\}\big\rangle&=\frac{V}{|\tilde{\mathbf{p}}|}\left(\frac{1}{2}+n_\mathbf{q}\right)\mathcal{P}_{ij} \ , \\
 \frac{1}{2}\big\langle\{{A}_{i,\mathbf{q}},{E}_{j,\mathbf{q}}\}\big\rangle&=0 \ , \\
 \frac{1}{2}\big\langle\{{E}_{i,\mathbf{q}},{E}_{j,\mathbf{q}}\}\big\rangle&=V|\tilde{\mathbf{p}}|\left(\frac{1}{2}+n_\mathbf{q}\right)\mathcal{P}_{ij} \ .
\end{align}
\end{subequations}
We note that vacuum initial conditions, i.e. quantum fluctuations around a coherent background field, are determined by $n_{\mathbf{q}}=0$.
\newline
\section{Fermionic observables}
\label{app:ferm_observables}
Given the mode function expansion of the Dirac field with vacuum initial conditions $n_{\lambda,\mathbf{q}}^{u}=n_{\lambda,\mathbf{q}}^{v}=0$, we can calculate the fermion energy density $\epsilon_{n}$ according to
\begin{align}
 \epsilon_{n}=-\frac{1}{2}\sum_{m\in\Lambda}\operatorname{tr}\{\mathcal{H}_{n,m}\Delta^{K}_{m,n}\} \ ,
\end{align}
where the trace is with respect to Dirac indices, and the lattice Hamiltonian including the spatial Wilson term is given by
\begin{widetext}
\begin{align}
 \mathcal{H}_{n,m}=\delta_{n_0,m_0}\left[\left(m+\sum\limits_{i}\frac{1}{a_i}\right)\delta_{\mathbf{n},\mathbf{m}}-\sum_{i}\frac{1}{2a_i}\left(i\gamma^i+1\right)U_{i,n}\delta_{\mathbf{n}+\hat{\imath},\mathbf{m}}+\sum_{i}\frac{1}{2a_i}\left(i\gamma^i-1\right)U_{-i,n}\delta_{\mathbf{n}-\hat{\imath},\mathbf{m}}\right] \ .
\end{align}
\end{widetext}
Expressed in terms of the mode functions, the energy density is
\begin{align}
 \epsilon_{n}=&\frac{1}{2V}\sum_{m\in\Lambda}\sum_{\mathbf{q}\in\tilde{\Lambda}}\sum_{\lambda} \nonumber \\
 &\left[\bar{\Phi}^v_{\lambda,n,\mathbf{q}}\mathcal{H}_{n,m}\Phi^{v}_{\lambda,m,\mathbf{q}}-\bar{\Phi}^u_{\lambda,n,\mathbf{q}}\mathcal{H}_{n,m}\Phi^{u}_{\lambda,m,\mathbf{q}}\right] \ .
\end{align}
In order to define the momentum distribution, we perform a discrete Fourier transformation of the mode functions:
\begin{align}
 \Phi^{u/v}_{\lambda,(m_0,\mathbf{m}),\mathbf{q}}\equiv\frac{1}{V}\sum_{\tilde{\mathbf{q}}\in\tilde{\Lambda}}{e^{i\tilde{\mathbf{p}}\cdot\mathbf{x}_m}{\Phi}^{u/v}_{\lambda,\tilde{\mathbf{q}},\mathbf{q}}} \ ,
\end{align}
with $\tilde{p}_i=2\pi\tilde{q}_i/N_ia_i$ for $i\in\{1,2,3\}$. 
Accordingly, this defines a discrete phase-space energy density:
\begin{align}
 \epsilon_{n,\tilde{\mathbf{q}}}=&\frac{1}{2V^2}\sum_{m\in\Lambda}e^{i\tilde{\mathbf{p}}\cdot\mathbf{x}_m}\sum_{\mathbf{q}\in\tilde{\Lambda}}\sum_{\lambda}\nonumber \\
 &\left[\bar{\Phi}^v_{\lambda,n,\mathbf{q}}\mathcal{H}_{n,m}{\Phi}^{v}_{\lambda,\tilde{\mathbf{q}},\mathbf{q}}-\bar{\Phi}^u_{\lambda,n,\mathbf{q}}\mathcal{H}_{n,m}{\Phi}^{u}_{\lambda,\tilde{\mathbf{q}},\mathbf{q}}\right] \ ,
\end{align}
such that
\begin{align}
 \epsilon_{n}=\sum_{\tilde{\mathbf{q}}\in\tilde{\Lambda}}\epsilon_{n,\tilde{\mathbf{q}}} \ .
\end{align}
We define the discrete phase-space particle number density as the total energy density divided by twice the single-particle energy density:
\begin{align}
 N_{n,\tilde{\mathbf{q}}}\equiv\frac{\epsilon_{n,\tilde{\mathbf{q}}}}{2\omega_{n,\tilde{\mathbf{q}}}} \ .
\end{align}
Here the single-particle energy density can be computed from the lattice Hamiltonian
\begin{align}
 \omega_{n,\tilde{\mathbf{q}}}=\sqrt{\bar{m}^2+\bar{p_1}^2+\bar{p_2}^2+\bar{p_3}^2} \ ,
\end{align}
with
\begin{subequations}
\begin{align}
 \bar{p}_i&=\frac{i}{2a_i}\left[U_{-i,n}e^{-\frac{2\pi i\tilde{q}_i}{N_i}}-U_{i,n}e^{\frac{2\pi i\tilde{q}_i}{N_i}}\right] \ , \\
 \bar{m}&=m+\sum_{i}\frac{1}{2a_i}\left[2-U_{i,n}e^{\frac{2\pi i\tilde{q}_i}{N_i}}-U_{-i,n}e^{-\frac{2\pi i\tilde{q}_i}{N_i}}\right] \ .
\end{align}
\end{subequations}
It can easily be checked that these expressions coincide with \eqref{eq:lat_mom} in the vacuum $U_{i,n}=1$. 
The normalized momentum distribution $N_{n_0,\tilde{\mathbf{q}}}$ -- in the main body of the text denoted as $n(p,t)$ -- is then given by:
\begin{align}
 n(p,t)\equiv N_{n_0,\tilde{\mathbf{q}}}=a_1a_2a_3\sum_{\mathbf{n}\in\Lambda}N_{n,\tilde{\mathbf{q}}} \ .
\end{align}
Similarly, we can also calculate the total fermion density $N_{n_0}$ -- in the main body of the text denoted as $N(t)/V$:
\begin{align}
 N(t)/V\equiv N_{n_0}=\frac{1}{N_1N_2N_3}\sum_{\mathbf{n}\in\Lambda}\sum_{\tilde{\mathbf{q}}\in\tilde{\Lambda}}N_{n,\tilde{\mathbf{q}}} \ .
\end{align}

\section{Continuum results for Schwinger effect}
\label{app:schwinger}
We briefly review some analytic results for the Schwinger effect in the static background field \cite{Hebenstreit:2010vz}.
In this case, the Dirac equation in the background field $E_0$ is analytically solvable in terms of parabolic cylinder functions $D_\nu(z)$.
Defining $\epsilon_0=gE_0/m^2$, $\epsilon_\perp^2=m^2+p_1^2+p_2^2$ such that $\omega^2(p)=\epsilon_\perp^2+p_3^2$ and $\eta=\epsilon_\perp^2/gE_0$, the analytic solution for the momentum distribution $f(p)$ yields:
\begin{align} 
 &f(p)=e^{-\pi\eta/4}\left[\frac{\eta}{2}\left(1-\frac{p_3}{\omega(p)}\right)\mathcal{D}_1(p)\right.\nonumber\\
 &\left.\quad+\left(1+\frac{p_3}{\omega(p)}\right)\mathcal{D}_2(p)-\sqrt{\frac{\epsilon_0\eta^2}{2}}\frac{m}{\omega(p)}\mathcal{D}_3(p)\right] \ , 
\end{align}
with
\begin{subequations}
\begin{eqnarray}
 \mathcal{D}_1(p)&=&\left|D_{-1+i\eta/2}(\hat{p})\right|^2 \ , \\
 \mathcal{D}_2(p)&=&\left|D_{i\eta/2}(\hat{p})\right|^2 \ , \\
 \mathcal{D}_3(p)&=&e^{i\pi/4}D_{i\eta/2}(\hat{p})D_{-1-i\eta/2}(\hat{p}^*) + c.c. \ , \qquad
\end{eqnarray}
\end{subequations}
for
\begin{align}
 \hat{p}&=-\sqrt{\frac{2}{\epsilon_0}}\frac{p_3}{m}e^{-i\pi/4} \ .
\end{align}
It can be shown that $f(p)$ vanishes for small kinetic momenta $p_3\to-\infty$ and approaches a non-vanishing constant for large kinetic momenta $p_3\to\infty$:
\begin{align}
 \lim_{p_3\to-\infty}f(p)=0 \quad , \quad \lim_{p_3\to\infty}f(p)=2e^{-\pi\eta} \ .
\end{align}Most notably, the rate at which electrons and positrons are created is a constant.
Accordingly, the total number $\Delta N$ of electrons and positrons, respectively, which are created per volume $V$ during a time interval $T$ is given by
\begin{align}
 \frac{\dot{N}}{V}=\frac{(gE_0)^2}{4\pi^3}\exp\left(-\frac{\pi m^2}{gE_0}\right)=\frac{m^4\epsilon_0^2}{4\pi^3}\exp\left(-\frac{\pi}{\epsilon_0}\right) \ .
\end{align}
\newline


\begin{thebibliography}{20}

\bibitem{Schwinger:1951nm}
  J.~S.~Schwinger,
  Phys.\ Rev.\  {\bf 82} (1951) 664.

\bibitem{Sauter:1931zz}
  F.~Sauter,
  Z.\ Phys.\  {\bf 69} (1931) 742.

\bibitem{Heisenberg:1935qt}
  W.~Heisenberg and H.~Euler,
  Z.\ Phys.\  {\bf 98} (1936) 714.

\bibitem{Gelis:2010nm}
  F.~Gelis, E.~Iancu, J.~Jalilian-Marian and R.~Venugopalan,
  Ann.\ Rev.\ Nucl.\ Part.\ Sci.\  {\bf 60} (2010) 463.

\bibitem{Berges:2010zv}
  J.~Berges, D.~Gelfand and J.~Pruschke,
  Phys.\ Rev.\ Lett.\  {\bf 107} (2011) 061301.
  
\bibitem{Son:1996zs}
  D.~T.~Son,
  hep-ph/9601377.

\bibitem{Khlebnikov:1996mc}
  S.~Y.~.Khlebnikov and I.~I.~Tkachev,
  Phys.\ Rev.\ Lett.\  {\bf 77} (1996) 219.
  
\bibitem{Prokopec:1996rr}
  T.~Prokopec and T.~G.~Roos,
  Phys.\ Rev.\ D {\bf 55} (1997) 3768.
   
\bibitem{Aarts:1998td}
  G.~Aarts and J.~Smit,
  Nucl.\ Phys.\ B {\bf 555} (1999) 355.
  
\bibitem{Gelis:2013oca}
 F.~Gelis and N.~Tanji,
 Phys.\ Rev.\ D {\bf 87} (2013) 125035.
  
\bibitem{Aarts:2001yn}
  G.~Aarts and J.~Berges,
  Phys.\ Rev.\ Lett.\  {\bf 88} (2002) 041603.

\bibitem{Polkovnikov:2003}
  A.~Polkovnikov,
  Phys.\ Rev.\ A {\bf 68} (2003) 053604.

\bibitem{Arrizabalaga:2004iw}
  A.~Arrizabalaga, J.~Smit and A.~Tranberg,
  JHEP {\bf 0410} (2004) 017.
  
\bibitem{Berges:2007ym}
  J.~Berges and T.~Gasenzer,
  Phys.\ Rev.\ A {\bf 76} (2007) 033604.

\bibitem{Berges:2008wm}
  J.~Berges, A.~.Rothkopf and J.~Schmidt,
  Phys.\ Rev.\ Lett.\  {\bf 101} (2008) 041603.
  
\bibitem{Berges:2013oba}
  J.~Berges, D.~Gelfand and D.~Sexty,
  Phys.\ Rev.\ D {\bf 89} (2014) 025001.
  
\bibitem{Berges:2013lsa}
  J.~Berges, K.~Boguslavski, S.~Schlichting and R.~Venugopalan,
  JHEP {\bf 1405} (2014) 054.
  
\bibitem{Epelbaum:2014yja}
  T.~Epelbaum, F.~Gelis and B.~Wu,
  arXiv:1402.0115 [hep-ph].   
   
\bibitem{Mueller:2002gd}
  A.~H.~Mueller and D.~T.~Son,
  Phys.\ Lett.\ B {\bf 582} (2004) 279.
  
\bibitem{Jeon:2004dh}
  S.~Jeon,
  Phys.\ Rev.\ C {\bf 72} (2005) 014907.
  
\bibitem{Mathieu:2014aba}
  V.~Mathieu, A.~H.~Mueller and D.~N.~Triantafyllopoulos,
  Eur.\ Phys.\ J.\ C {\bf 74} (2014) 2873.
  
\bibitem{Berges:2013eia}
  J.~Berges, K.~Boguslavski, S.~Schlichting and R.~Venugopalan,
  Phys.\ Rev.\ D {\bf 89} (2014) 074011.
 
  
\bibitem{Berges:2013fga}
  J.~Berges, K.~Boguslavski, S.~Schlichting and R.~Venugopalan,
  Phys.\ Rev.\ D {\bf 89} (2014) 114007

\bibitem{York:2014wja}
  M.~C.~A.~York, A.~Kurkela, E.~Lu and G.~D.~Moore,
  Phys.\ Rev.\ D {\bf 89} (2014) 074036.
 
\bibitem{Hebenstreit:2013qxa}
  F.~Hebenstreit, J.~Berges and D.~Gelfand,
  Phys.\ Rev.\ D {\bf 87} (2013) 105006.
 
\bibitem{Schwinger:1961}
  J.~S.~Schwinger,
  J.\ Math.\ Phys.\  {\bf 2} (1961) 407.

\bibitem{Keldysh:1965}
  L.~V.~Keldysh,
  Zh.\ Eksp.\ Teor.\ Fiz.\  {\bf 47} (1964) 1515 [Sov.\ Phys.\ JETP {\bf 20} (1965) 1018].
 
\bibitem{Ambjorn:1990pu}
  J.~Ambjorn, T.~Askgaard, H.~Porter and M.~E.~Shaposhnikov,
  Nucl.\ Phys.\ B {\bf 353} (1991) 346.

\bibitem{Berges:2007re}
  J.~Berges, S.~Scheffler and D.~Sexty,
  Phys.\ Rev.\ D {\bf 77} (2008) 034504.

\bibitem{Berges:2004yj}
  J.~Berges,
  AIP Conf.\ Proc.\  {\bf 739} (2005) 3;
  hep-ph/0409233.

\bibitem{Kawai:1980ja}
  H.~Kawai, R.~Nakayama and K.~Seo,
  Nucl.\ Phys.\ B {\bf 189} (1981) 40.
  
\bibitem{Danielewicz:1982ca}
  P.~Danielewicz,
  Annals Phys.\  {\bf 152} (1984) 305.

\bibitem{Furry:1937zz}
  W.~H.~Furry,
  Phys.\ Rev.\  {\bf 51} (1937) 125.
  
\bibitem{Nielsen:1981}
  H.~B.~Nielsen and M.~Ninomiya,
  Phys.\ Lett.\ B {\bf 105} (1981) 219.
  
\bibitem{Wilson:1974sk}
  K.~G.~Wilson,
  Phys.\ Rev.\ D {\bf 10} (1974) 2445.
  
\bibitem{Kogut:1975}
  J.~B.~Kogut and L.~Susskind,
  Phys.\ Rev.\ D {\bf 11} (1975) 395.

\bibitem{Kaplan:1992bt}
  D.~B.~Kaplan,
  Phys.\ Lett.\ B {\bf 288} (1992) 342.

\bibitem{Frezzotti:2000nk}
  R.~Frezzotti, P.~A.~Grassi, S.~Sint and P.~Weisz [Alpha Collaboration],
  JHEP {\bf 0108} (2001) 058.
  
\bibitem{Borsanyi:2008eu}
  S.~Borsanyi and M.~Hindmarsh,
  Phys.\ Rev.\ D {\bf 79} (2009) 065010.

\bibitem{Mou:2013kca}
  Z.~-G.~Mou, P.~M.~Saffin and A.~Tranberg,
  JHEP {\bf 1311} (2013) 097.
  
\bibitem{Cooper:1996ii}
  F.~Cooper, S.~Habib, Y.~Kluger and E.~Mottola,
  Phys.\ Rev.\ D {\bf 55} (1997) 6471.

\bibitem{Leibbrandt:1987qv}
  G.~Leibbrandt,
  Rev.\ Mod.\ Phys.\  {\bf 59} (1987) 1067.
  
\bibitem{Kluger:1992gb}
  Y.~Kluger, J.~M.~Eisenberg, B.~Svetitsky, F.~Cooper and E.~Mottola,
  Phys.\ Rev.\ D {\bf 45} (1992) 4659.
    
\bibitem{Bloch:1999eu}
  J.~C.~R.~Bloch, V.~A.~Mizerny, A.~V.~Prozorkevich, C.~D.~Roberts, S.~M.~Schmidt, S.~A.~Smolyansky and D.~V.~Vinnik,
  Phys.\ Rev.\ D {\bf 60} (1999) 116011.
  
\bibitem{BialynickiBirula:1991tx}
  I.~Bialynicki-Birula, P.~Gornicki and J.~Rafelski,
  Phys.\ Rev.\ D {\bf 44} (1991) 1825.
  
\bibitem{Hebenstreit:2011wk}
  F.~Hebenstreit, R.~Alkofer and H.~Gies,
  Phys.\ Rev.\ Lett.\  {\bf 107} (2011) 180403

\bibitem{Berenyi:2013eia}
  D.~Berényi, S.~Varró, V.~V.~Skokov and P.~Lévai,
  [arXiv:1401.0039 [hep-ph]].
  
\bibitem{Kobes:1985}
  R.~L.~Kobes, G.~W.~Semenoff and N.~Weiss,
  Z.\ Phys.\ C {\bf 29} (1985) 371.

\bibitem{Cooper:1994hr}
  F.~Cooper, S.~Habib, Y.~Kluger, E.~Mottola, J.~P.~Paz and P.~R.~Anderson,
  Phys.\ Rev.\ D {\bf 50} (1994) 2848.

\bibitem{Rothe:1992nt}
  H.~J.~Rothe,
  World Sci.\ Lect.\ Notes Phys.\  {\bf 82} (2012) 1.
     
\bibitem{Hebenstreit:2010vz}
  F.~Hebenstreit, R.~Alkofer and H.~Gies,
  Phys.\ Rev.\ D {\bf 82} (2010) 105026.

\end{thebibliography}
\end{document}